\documentclass[dvips]{acta}
\usepackage{supertabular,lscape,epsfig}
\usepackage{amssymb}
\usepackage{amsmath}
\usepackage{upgreek}
\usepackage{graphicx}
\usepackage{longtable}
\usepackage{pdflscape}

\SetPages{0}{0}

\SetVol{58}{2020}

\begin{document}

\begin{Titlepage}
\Title{On the pulsations of the $\delta$~Scuti star of the binary system KIC~6629588}
\Author{A. L~i~a~k~o~s}{Institute for Astronomy, Astrophysics, Space Applications and Remote Sensing, National Observatory of Athens, Metaxa \& Vas. Pavlou St., GR-15236, Penteli, Athens, Greece\\
e-mail: alliakos@noa.gr}

\Received{Month Day, Year}
\end{Titlepage}

\Abstract{This work includes a comprehensive analysis of the $Kepler$ detached eclipsing binary system KIC~6629588 that aims to the detailed study of the oscillation properties of its pulsating component. Ground-based spectroscopic observations were obtained and used to classify the components of the system. The spectroscopic results were used as constrains for the modelling of the short-cadence $Kepler$ light curves and for the estimation of the absolute parameters of the components. Furthermore, the light curves residuals are analyzed using Fourier transformation techniques in order to search for pulsation frequencies. The primary component of the system is identified as a $\delta$~Scuti star that pulsates in seven eigenfrequencies in the range 13-22~d$^{-1}$, while more than 270 combination frequencies were also detected. 
The absolute and the oscillation parameters of this pulsating star are compared with those of other $\delta$~Scuti stars-members of detached binary systems using evolutionary and correlation diagrams. Finally, the distance of the system is also estimated.}
{stars:binaries:eclipsing -- stars:fundamental parameters -- (Stars:) binaries (including multiple): close -- Stars: oscillations (including pulsations) -- Stars: variables: delta Scuti -- Stars: individual: KIC~6629588}

\section{Introduction}

The $\delta$~Scuti stars are short-period pulsating variables that may oscillate in both radial and non-radial modes. The radial and the low-order non-radial modes are generated due to $\kappa$-mechanism (\textit{c.f.} Aerts \textit{et al.} 2010, Balona \textit{et al.} 2015), while the higher-order non-radial ones are caused due to the turbulent pressure in the Hydrogen convective zone (Antoci \textit{et al.} 2014, Grassitelli \textit{et al.} 2015). They have typical masses 1.4-2.5~$M_{\odot}$, temperatures approximately between 6500-9500~K, luminosity classes III-V, and their majority lie within the classical instability strip (Aerts \textit{et al.} 2010, Murphy \textit{et al.} 2019).
The eclipsing binary systems (EBs) are outstanding tools for the calculation of the absolute parameters (\textit{e.g.} masses, radii, luminosities) of their components, that leads to the estimation of their evolutionary status. Therefore, the EBs hosting a pulsating component can be considered as very special cases, since they provide valuable knowledge about the properties and evolution of pulsating stars. Moreover, these cases offer the unique opportunity to collect information about the affect of binarity (\textit{e.g.} proximity, mass transfer) on the evolution of the pulsations.

Mkrtichian \textit{et al.} (2002) proposed the term `$oEA~stars$' (oscillating eclipsing binaries of Algol type) for the EBs with a $\delta$~Scuti mass accretor component of (B)A-F spectral type. The first correlation between orbital ($P_{\rm orb}$) and dominant pulsation ($P_{\rm pul}$) periods for these systems was published by Soydugan \textit{et al.} (2006), followed by updated ones by Liakos \textit{et al.} (2012), Zhang \textit{et al.} (2013), and Liakos and Niarchos (2017). Liakos and Niarchos (2015, 2016, 2017) noticed that for systems with $P_{\rm orb}>$13~d, these quantities are uncorrelated. This threshold was re-estimated as approximately 25~d by Kahraman Ali\c{c}avu\c{s} \textit{et al.} (2017), based only on EBs. The most complete catalogue for binaries with a $\delta$~Scuti component was published by Liakos and Niarchos (2017), it is available online (http://alexiosliakos.weebly.com/catalogue.html), and it is frequently updated. The latter work includes also correlations between the fundamental parameters of these systems according to their geometrical status. Murphy (2018) published a review on this kind of systems. Murphy \textit{et al.} (2018), using approximately 2224 $\delta$~Scuti light curves, observed by the $Kepler$ mission, identified indirectly 341 new binaries with a $\delta$~Scuti member that have relatively long $P_{\rm orb}$ ($>20$~d). Liakos (2020) published updated correlations between $P_{\rm pul}-P_{\rm orb}$ and $P_{\rm pul}-\log g$ specifically for detached eclipsing binaries with a $\delta$~Scuti component with $P_{\rm orb}<13$~d.

The utmost accuracy ($10^{-4}$~mag) and the continuity of the data obtained by the $Kepler$ (Borucki \textit{et al.} 2010, Koch \textit{et al.} 2010) and $K2$ (Howell \textit{et al.} 2014) missions provide the basis for highly detailed asteroseismic studies, especially when these data are combined with multi-colour photometry or high-resolution spectroscopy. Especially the short-cadence data (time resolution $\sim$1~min) are considered as extremely useful for the modelling of high-frequency pulsators (\textit{e.g.} $\delta$~Scuti stars) and allow the detection of oscillations with amplitudes up to a few $\upmu$mag (\textit{c.f.} Murphy \textit{et al.} 2013, Liakos 2017, Bowman and Kurtz 2018, Lee \textit{et al.} 2019b, Kurtz \textit{et al.} 2020). A very useful database for EBs observed by the $Kepler$ mission, namely the `$Kepler$ Eclipsing Binary Catalog' (KEBC-http://keplerebs.villanova.edu/), has been established by Pr\v{s}a \textit{et al.} (2011) and provides tons of detrended data for a few thousands of systems.

This work is part of the paper-series on individual EBs with a $\delta$~Scuti component (see also Liakos and Niarchos 2009, 2013, 2016, 2020, Soydugan \textit{et al.} 2013, Liakos and Caga\v{s} 2014, Liakos \textit{et al.} 2008, 2012, Liakos 2017, 2018, 2020, Ula\c{s} \textit{et al.} 2020). It aims to present in detail the pulsational and physical properties of the oscillating member of KIC~6629588 and, more general, to contribute in the topic of asteroseismology of close detached binary systems. According to Liakos (2020) and Liakos and Niarchos (2020), there are only 44 known systems of this type with $P_{\rm orb}<13$~d, while only 16 of them have been observed by space missions (\textit{i.e.} with detailed asteroseismic modelling). Therefore, the present paper can be considered as an appreciable contribution to the specific study of close detached EBs with $\delta$~Scuti members, whose oscillation properties are potentially influenced by the binarity.

KIC~6629588 (2MASS~J19510135+4200326) was identified as EB by the $Kepler$ mission (Slawson \textit{et al.} 2011) and has a period of $\sim2.26$~d. Various estimations for its temperature have been made (Slawson \textit{et al.} 2011, Pinsonneault \textit{et al.} 2012, Huber \textit{et al.} 2014, Frasca \textit{et al.} 2016) and range between 6500-7130~K. Liakos and Niarchos (2016) published preliminary results for the system and derived the dominant frequency of the pulsating component as $\sim13.4$~d$^{-1}$.

The ground-based spectroscopic observations and their results are presented in Section~2. Section~3 hosts the methods and the results for the light curves (LC) modelling and the calculation of the physical parameters of the system. The pulsation analysis 
of the $\delta$~Scuti component are given in Section~4. Finally, the discussion on the evolution of the system, the comparison of the pulsator's properties with other similar cases, the estimation of the distance of the system, and the conclusions of the work are presented in Section~5.


\section{Spectroscopic observations}

The spectroscopic observations were aiming to classify the component(s) of the system. The data were obtained with the 2.3~m Ritchey-Cretien `Aristarchos' telescope at Helmos Observatory in Greece in August 2016. The \emph{Aristarchos~Transient~Spectrometer} instrument (http://helmos.astro.noa.gr/ats.html; Boumis \textit{et al.} 2004) using the low resolution grating (600~lines~mm$^{-1}$) was employed for the observations. This set-up provided a resolution of $\sim3.2$~{\AA}~pixel$^{-1}$ and a spectral coverage between approximately 4000-7260~{\AA}. Two successive spectra with 15~min exposures were acquired for KIC~6629588 at the system's orbital phase 0.79 and added together in order to achieve a better signal-to-noise ratio. In order to classify the component(s) of the system, standard stars between A0-K8 spectral classes (one standard star per subclass) were also observed with exactly the same set-up between August-October 2016. All spectra were calibrated (bias, dark, flat-field corrections) using the \textsc{MaxIm DL} software. The data reduction (wavelength calibration, cosmic rays removal, spectra normalization, sky background removal) was made with the \textsc{RaVeRe} v.2.2c software (Nelson 2009).

For the classification, a spectral correlation technique and a spectral disentangling method were applied (for details see Liakos 2017, 2020). However, given that in the spectrum of KIC~6629588 a noticeable light contribution from the secondary component was detected, the aforementioned spectral disentangling method was used. Briefly, this method uses the spectra of the standard stars to calculate combined spectra with different light contributions from the components (the starting value for the contribution of the primary component is 0.5 and the step is 0.05). Each combined spectrum is compared in terms of sums of squared residuals ($\Sigma res^2$) with the spectrum of the studied system, based on the Balmer and some strong metallic lines. The least value of these $\Sigma res^2$ indicates the best match between the spectra.

According to this method, it was found that the spectrum of KIC~6629588 fits very well to a combined spectrum of F1V+K4V spectral type standard stars, with an 85\% luminosity contribution of the primary ($\Sigma res^2=0.32$). The $\Sigma res^2$ against all possible spectra combinations and their sub-combinations (\textit{i.e.} different light contributions of the components) are plotted in Fig.~1. The spectrum of the system along with the best-match standard stars combination spectrum is illustrated in Fig.~2. The spectral classification resulted in a spectral type F1V for the primary and K4V for the secondary component, with an error assumption of one sub-class. Hence, the corresponding temperatures ($T_{\rm eff}$), based on the relations between $T_{\rm eff}$ and spectral types of Cox (2000), are $7150\pm150$~K and $4500\pm140$~K for the primary and secondary components, respectively. The present results come in relatively good agreement with those given in previous studies (see Section~1). It should to be noted that the second best-match between the observed spectrum and the combined spectrum was found for the combination of A9V+G3V spectral type standard stars ($\Sigma res^2=0.35$), with a 65\% luminosity contribution of the primary. This solution was also tested in the subsequent analysis.

\begin{figure}[h]
\begin{center}
\includegraphics[width=12cm]{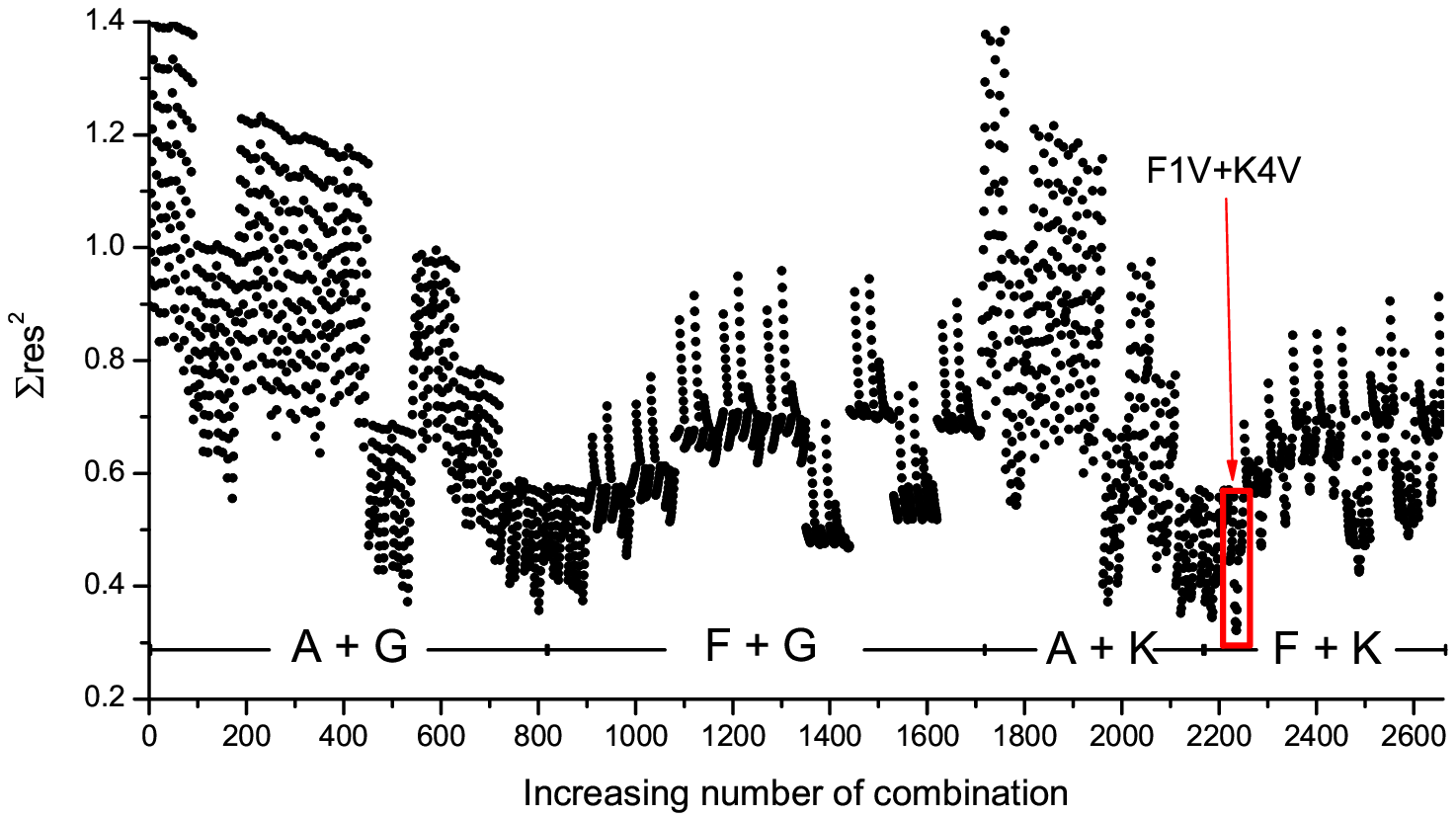}
\end{center}
	\FigCap{Spectral type-search plot for KIC~6629588 based on the spectra combination method. The combinations of the spectra are indicated along the horizontal axis. Each combination consists of ten individual sub-combinations with different luminosity contributions of the components (for details see Liakos 2020). The best match is found for the combination of F1V+K4V spectral types with an 85\% luminosity contribution of the primary.}
\end{figure}
\begin{figure}[h]
\begin{center}
\includegraphics[width=12cm]{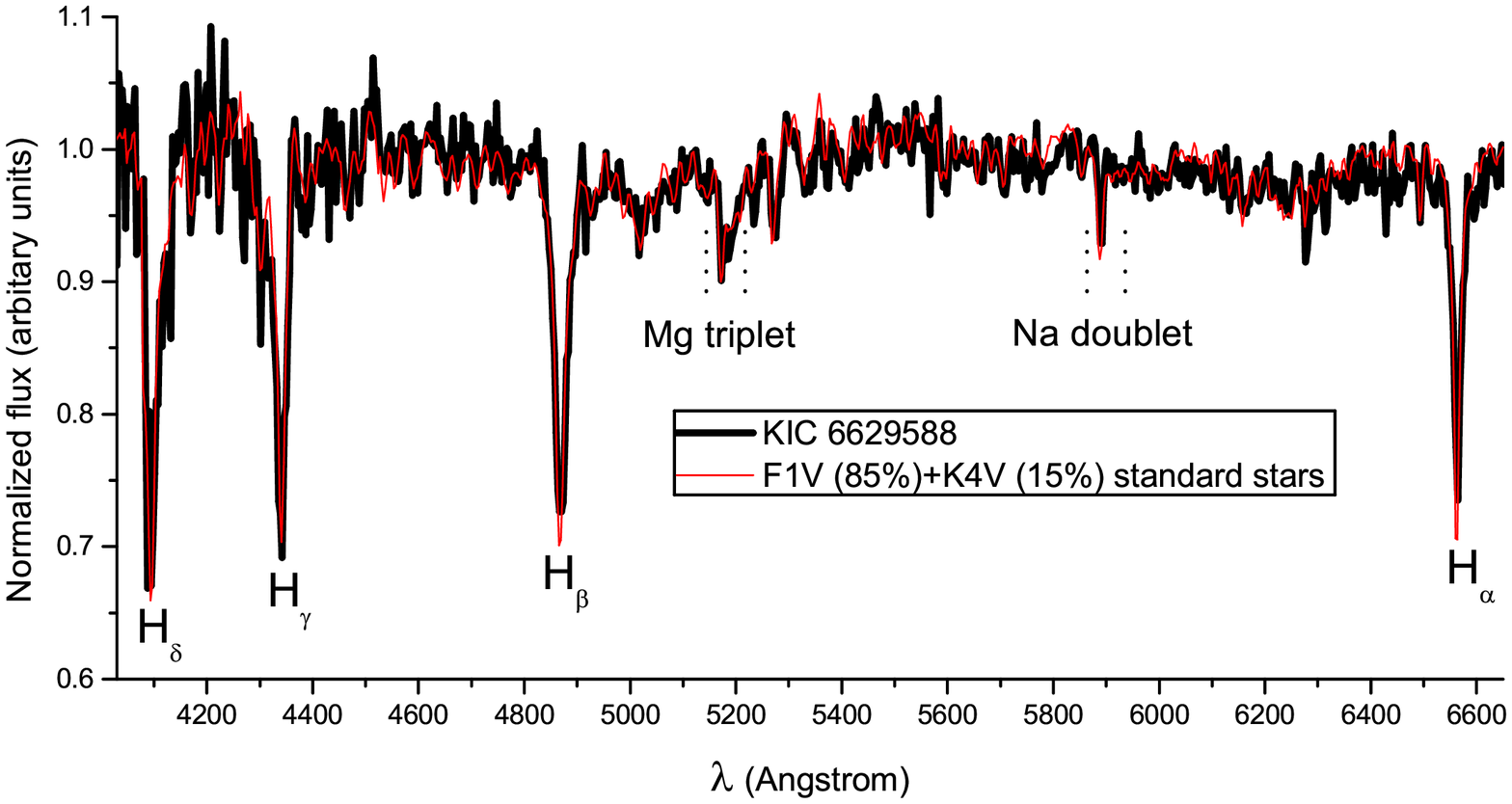}
\end{center}
		\FigCap{Comparison between the spectrum of KIC~6629588 (black lines) and a combination spectrum (red lines) of standard stars of F1V (85\% contribution) and K4V (15\% contribution) spectral types. The Balmer and some strong metallic lines are also indicated.}
\end{figure}

\newpage

\section{Light curve modelling and calculation of physical parameters}

KIC~6629588 was observed in both long- and short-cadence modes by the $Kepler$ mission. Nevertheless, given that the primary aim of the present work is the estimation of the pulsational properties of the oscillating component, only the short-cadence data, due to their much better time resolution (\textit{i.e.} $\sim$1~min against $\sim$30~min of the long-cadence data), were used for the subsequent analyses. The system was observed only during one quarter (Q11) in short-cadence mode, providing in total 134803 data points in a time span of 97.1 successive days that yield in 41 fully covered (orbital cycles) LCs. 
A small sample of these $Kepler$ LCs is illustrated in Fig.~3. Using the ephemeris $Min.~I=BJD~2454966.78(4)+ 2.264471(3)\times E$ and the $Kepler$ magnitude $K_{\rm p}$=13.98~mag of the system (taken from $KEBC$), the orbital phases and the magnitudes of the sample data points were respectively calculated.

The subsequent LCs analyses employed the software \textsc{PHOEBE} v.0.29d (Pr\v{s}a and Zwitter 2005). The temperatures of the components were given initially values as derived from spectroscopy (see Section~2). However, given the higher certainty in the determination of the spectral type of the primary component due to its greater light contribution in the total light ($\sim$85\%), the $T_{\rm eff}$ of this component was kept fixed, while that of the secondary was adjusted. It should to be noted that the second best-match solution from spectroscopy (\textit{i.e.} A9V+G3V) was also tested. However, the latter temperature difference, that is reflected to the brightness difference, hence the minimum light differences, could not be reproduced by any model in the light curve analysis. Therefore, the temperatures of the components as derived from the best fit combined spectrum (\textit{i.e.} F1V+K4V) were used, as described before, in the subsequent modelling. According to the present spectroscopic results, radiative and convective envelopes were assumed for the primary and secondary components, respectively. Therefore, the corresponding values for the albedos ($A$) and the gravity darkening coefficients ($g$) as given in Ruci\'{n}ski (1969), von Zeipel (1924), and Lucy (1967) were adopted. The (linear) limb darkening coefficients ($x$) were taken from the lists of van Hamme (1993). The dimensionless potentials ($\Omega$), the fractional luminosity of the primary component ($L_{1}$), and the inclination of the system ($i$) were set as adjustable parameters. 
Regarding the filter depended parameters ($L_1$ and $x$), the filter $R_{\rm c}$ was chosen as the closest one to the spectral range covered by the cameras of $Kepler$ (\textit{c.f.} Liakos 2017, 2020).

\begin{figure}
\includegraphics[width=\columnwidth]{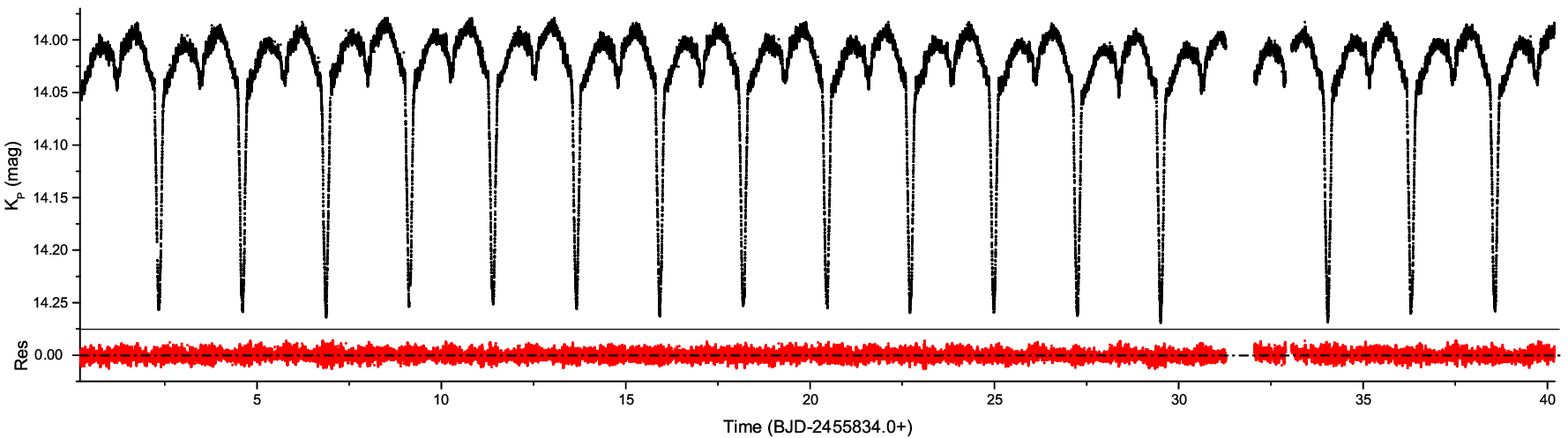}\\
\FigCap{Short-cadence LCs (black points) for KIC~6629588 and their residuals (red points) after the subtraction of the LC models. The plotted data concern only the 40 out of 97.1 days of observations, while the rest are not shown due to scaling reasons.}
\label{fig:LCsandRes}
\end{figure}

Due to the lack of radial velocity curves for this system, the mass ratio ($q$)-search method (for details see \textit{e.g.} Liakos and Niarchos 2012) was applied to estimate the most possible $q$ (\textit{c.f.} Soydugan and Ka\c{c}ar 2013, Liakos 2017, Wang \textit{et al.} 2017, Zhang \textit{et al.} 2017, Samadi Ghadim \textit{et al.} 2018, Yang \textit{et al.} 2018, Lee \textit{et al.} 2019a, 2019b, 2020). For the $q$-search, a mean LC was used in order to exclude as much as possible the pulsations from the modelling. This mean LC was the average of three LCs. It should be noted that a complete LC of this system contains approximately 3300 data points, while the mean LC, using averaged points per phase, includes approximately 300 normal points. Hence, using this method, the variations of the pulsations are vanished and do not affect at all the modelling. The $q$-search was applied in modes 2 (detached system), 4 (semi-detached system with the primary component filling its Roche lobe) and 5 (semi-detached system with the secondary component filling its Roche lobe). The step of $q$ change during the search was 0.1 starting from $q$=0.1. The sums of the squared residuals were systematically lower for all $q$ values in mode 2, therefore, it can be plausibly considered as a detached EB.

The $q$-search method for this system resulted in $q$=0.8 as shown in Fig.~4. This value was set as initial parameter in the normal LCs analyses (\textit{i.e.} including all points) and was left free to converge. Moreover, due to brightness asymmetries in the quadratures, a cool spot on the surface of the secondary star was also assumed and its parameters (latitude $Co-lat$, longitude $long$, radius, and temperature factor $Tf$) were adjusted. The selection of the spotted component was based on the temperatures of the members of this EB, thus, the secondary was found to fit better to a profile of a magnetically active star. Using the method described in Liakos (2017, 2018, 2020; \textit{i.e.} one model per group of successive LCs), 14 individual LC models were obtained and averaged to create the final model. Using this method, the final averaged model has more realistic errors estimations and the LC residuals are exempted of the spots effects, that is quite important for the following frequency search (Section~4).

The averaged photometric model is given in Table~1, while an example of LC fitting and the 2-D Roche geometry of the system are illustrated in Fig.~5. Part of the LCs residuals, after the subtraction of the individual models, is shown in the Fig.~3. The parameters of the spot are given in Appendix~A (Table~3). The corresponding immigration plot as well as the location of the spot on the secondary component are shown in Fig.~12 in the same appendix.

\begin{figure}
	\begin{center}
\includegraphics[width=10cm]{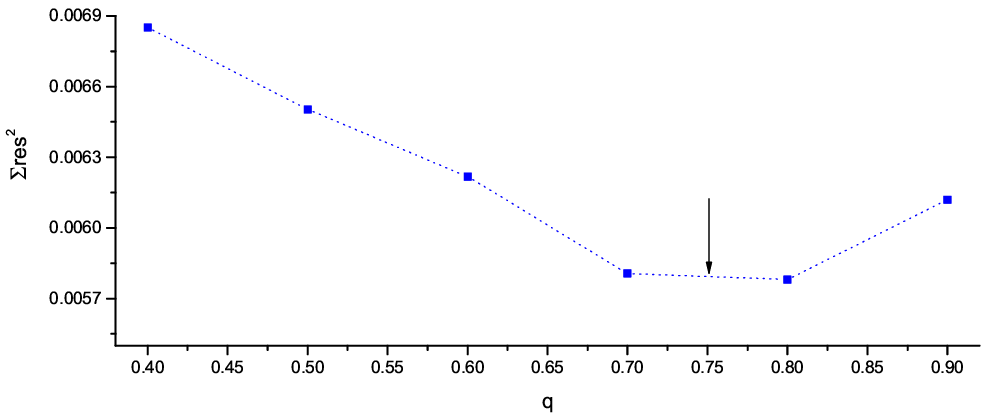}
\end{center}
	\FigCap{$q$-search plot for KIC~6629588. The arrow denotes the final $q$ value after the iterations.}
\end{figure}

\begin{figure}
\begin{center}
\includegraphics[width=10cm]{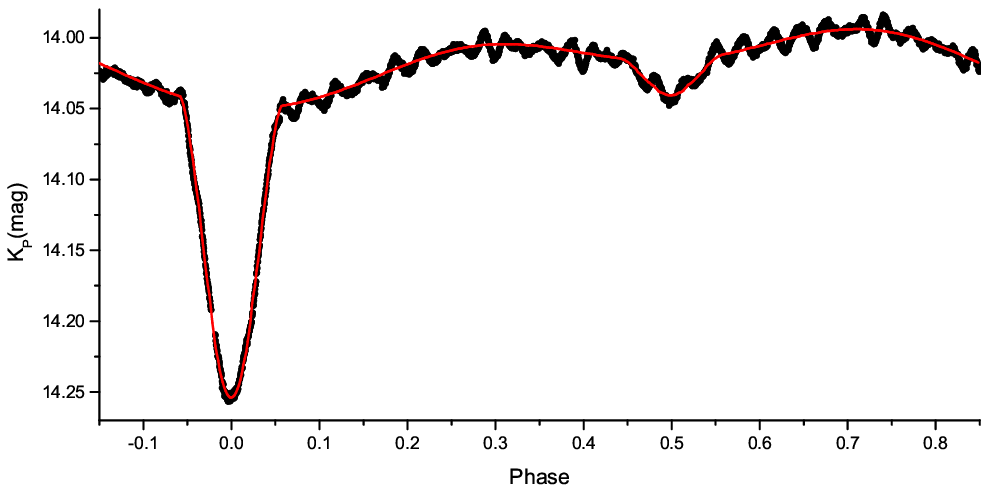}\\
\includegraphics[width=5cm]{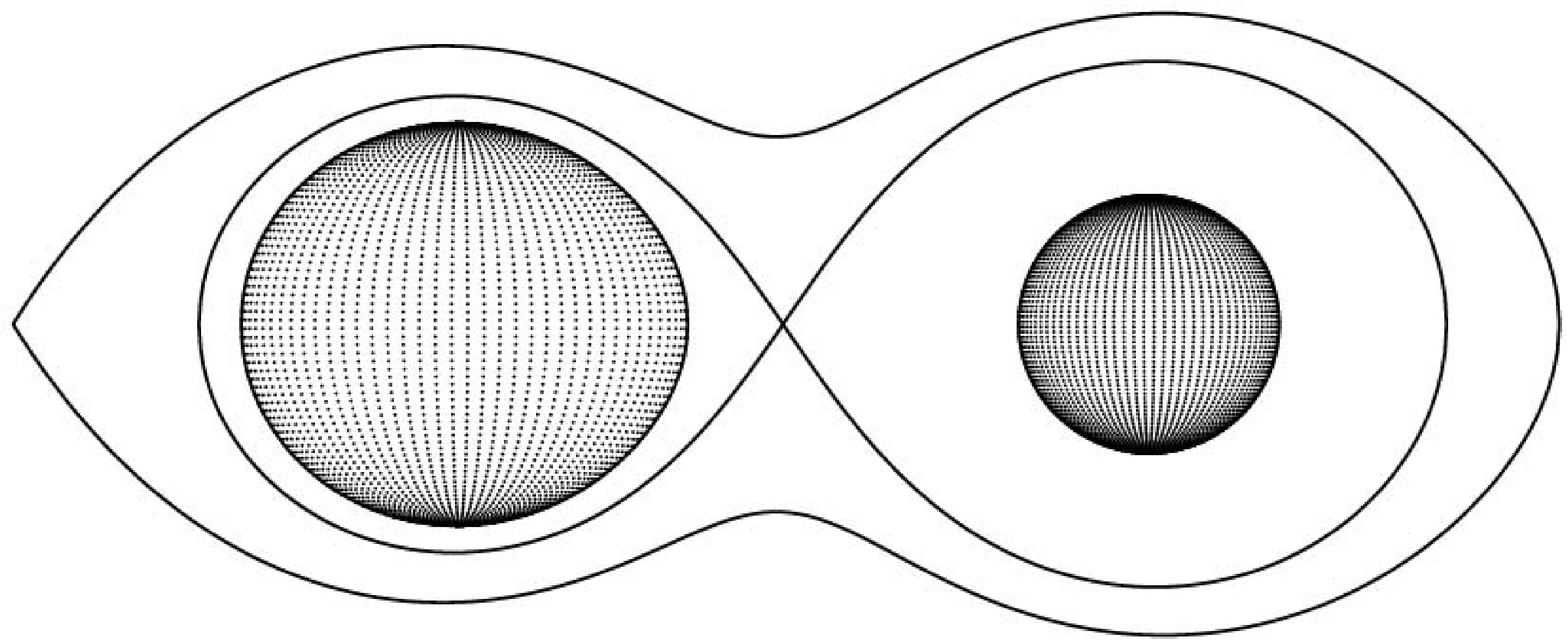}
\end{center}
	\FigCap{Synthetic (solid line) LCs over the observed $Kepler$ LCs (points) of an individual orbital cycle and the Roche geometry plot at orbital phase 0.75.}
\end{figure}

Based on the photometric calculation of $q$ and on a plausible assumption about the mass of the primary component, the absolute parameters of the system can be fairly estimated. According to the results of Liakos and Niarchos (2017) regarding the main-sequence nature of the majority of the $\delta$~Scuti stars in binaries and using the spectral type-mass correlations of Cox (2000) for main-sequence stars, a mass value of 1.67($\pm10\%$)~$M_{\odot}$ for the primary component can be adopted. Subsequently, the mass of the secondary component yields from the calculated mass ratio, followed by the derivation of the semi-major axes $a$ based on the Kepler's third law. The luminosities ($L$), the gravity acceleration ($\log g$), and the bolometric magnitudes ($M_{\rm bol}$) were derived according to the standard definitions. All the aforementioned absolute parameters were calculated using the software \textsc{AbsParEB} (Liakos 2015) and they are listed in Table~1.

	\MakeTable{lcc lcc}{13cm}{Photometric modelling results and absolute parameters for KIC~6629588. The errors are given in parentheses alongside values and correspond to the last digit(s).}
{\hline											
	&	\multicolumn{2}{c}{Components parameters}			&		&	\multicolumn{2}{c}{Absolute parameters}			\\
\hline											
	&	Primary	&	Secondary	&		&	Primary	&	Secondary	\\
\hline											
$T_{\rm eff}$~(K)	&	7150(150)$^{(1)}$	&	4405(110)	&	$M~$($M_{\odot}$)	&	1.56(16)$^a$	&	1.16(2)	\\
$\Omega$	&	6.06(7)	&	3.64(1)	&	$R~$($R_{\odot}$)	&	1.96(3)	&	3.06(3)	\\
$A$	&	1$^a$	&	0.5$^a$	&	$L~$($L_{\odot}$)	&	9.0(8)	&	3.2(4)	\\
$g$	&	1$^a$	&	0.32$^a$	&	$\log g$~(cm~s$^{-2}$)	&	4.05(2)	&	3.53(1)	\\
$x$	&	0.410	&	0.695	&	$a$~($R_{\odot}$)	&	4.43(4)	&	5.94(2)	\\
$L_i$/$L_{\rm T}$	&	0.737(9)	&	0.263(2)	&	$M_{\rm bol}$~(mag)	&	2.36(5)	&	3.50(8)	\\
\cline{4-6}											
$r_{\rm pole}$	&	0.188(1)	&	0.292(1)	&		&	\multicolumn{2}{c}{System parameters}			\\
\cline{4-6}											
$r_{\rm point}$	&	0.190(1)	&	0.331(2)	&	$q$~($m_{2}$/$m_{1}$)	&		\multicolumn{2}{c}{0.75(1)}		\\
$r_{\rm side}$	&	0.189(1)	&	0.301(1)	&	$i$~($\deg$)	&		\multicolumn{2}{c}{69.0(1)}		\\
$r_{\rm back}$	&	0.190(1)	&	0.317(2)	&	$P_{\rm orb}^{a,b}$~(d)	& \multicolumn{2}{c}{2.264471(3)}				\\
\hline											
	\multicolumn{6}{p{12cm}}{$^1$Section~2, $^a$assumed, $^bKEBC$, $i$=1 or 2, $L_T=L_1+L_2$}
}

\newpage
\section{Pulsations frequencies search}
The search for pulsations employed the software \textsc{PERIOD04} v.1.2 (Lenz and Breger 2005) that is based on classical Fourier analysis and was made on the LCs residuals of the system (Fig.~3) in the range 0-80~d$^{-1}$. The search range was selected in order to: a) cover the typical frequency ($f$) range of $\delta$~Scuti stars (\textit{i.e.} 4-80~d$^{-1}$; Breger 2000) and b) include the regime 0-4~d$^{-1}$, where longer-period oscillations may be exhibited either due to $g$-mode pulsations (\textit{i.e.} hybrid behaviour of $\gamma$~Doradus-$\delta$~Scuti type; \textit{e.g.} Uytterhoeven \textit{et al.} 2011, Zhang \textit{et al.} 2019) or due to combinations of higher frequencies (\textit{e.g.} Liakos 2017, 2020). The data points of the eclipses (\textit{i.e.} between 0.94-0.06 and 0.44-0.56 orbital phases) were excluded from the analysis because during these phaseparts the total light varies due to geometric reasons, hence, the amplitudes of the pulsations are affected. 
The recommended limit of the software for a frequency detection is the signal-to-noise ratio (S/N)=4 (\textit{i.e.} the amplitude of the detected frequency should be four times greater than the local background noise). However, according to Baran \textit{et al.} (2015), for shorter time series (order of a few months), a S/N$>$5 is recommended as reliability limit of a detected frequency, hence, this limit is also adopted for the present study. Nevertheless, given the complex nature of $\delta$~Scuti stars regarding the density of their frequencies, the calculated S/N from the software is not always trustable. Therefore, as described in Liakos (2017, 2018), the background noise was calculated in regions where no frequencies seem to exist (\textit{e.g.} 60-70~d$^{-1}$) with a spacing of 2~d$^{-1}$ and a box size of 2. This method resulted in a background noise of 8.58~$\upmu$mag that corresponds to a 5-$\sigma$ value of 0.043~mmag. After the first frequency computation the residuals were subsequently pre-whitened for the next one until the detected frequency had an amplitude with S/N$>$5. The Nyquist frequency of the data sample is 244.6~d$^{-1}$ and the frequency resolution, according to Loumos and Deeming (1978) (\textit{i.e.} 1.5/$T$, where $T$ is the observations time range in days), is 0.015~d$^{-1}$. 


The relation of Breger (2000), that correlates the pulsation constant $Q$ with the $T_{\rm eff}$, $\log g$, $M_{\rm bol}$, and $f$, was used to calculate the $Q$ value of each independent frequency. Moreover, given that the dominant pulsation frequency varies from time to time in $\delta$~Scuti stars, the average $Q$ value of the independent frequencies was subsequently used in the pulsation constant-density relation ($Q-\rho_{\rm pul}$), in order to estimate the density of the pulsator $\rho_{\rm pul}$ (\textit{c.f.} Liakos 2020). The ratio $P_{\rm pul}$/$P_{\rm orb}$ of all independent frequencies was calculated and found below 0.07, that is the upper value for the $p$-mode oscillations (Zhang \textit{et al.} 2013).

\begin{figure}
	\centering
	\begin{tabular}{cc}
	\includegraphics[width=11cm]{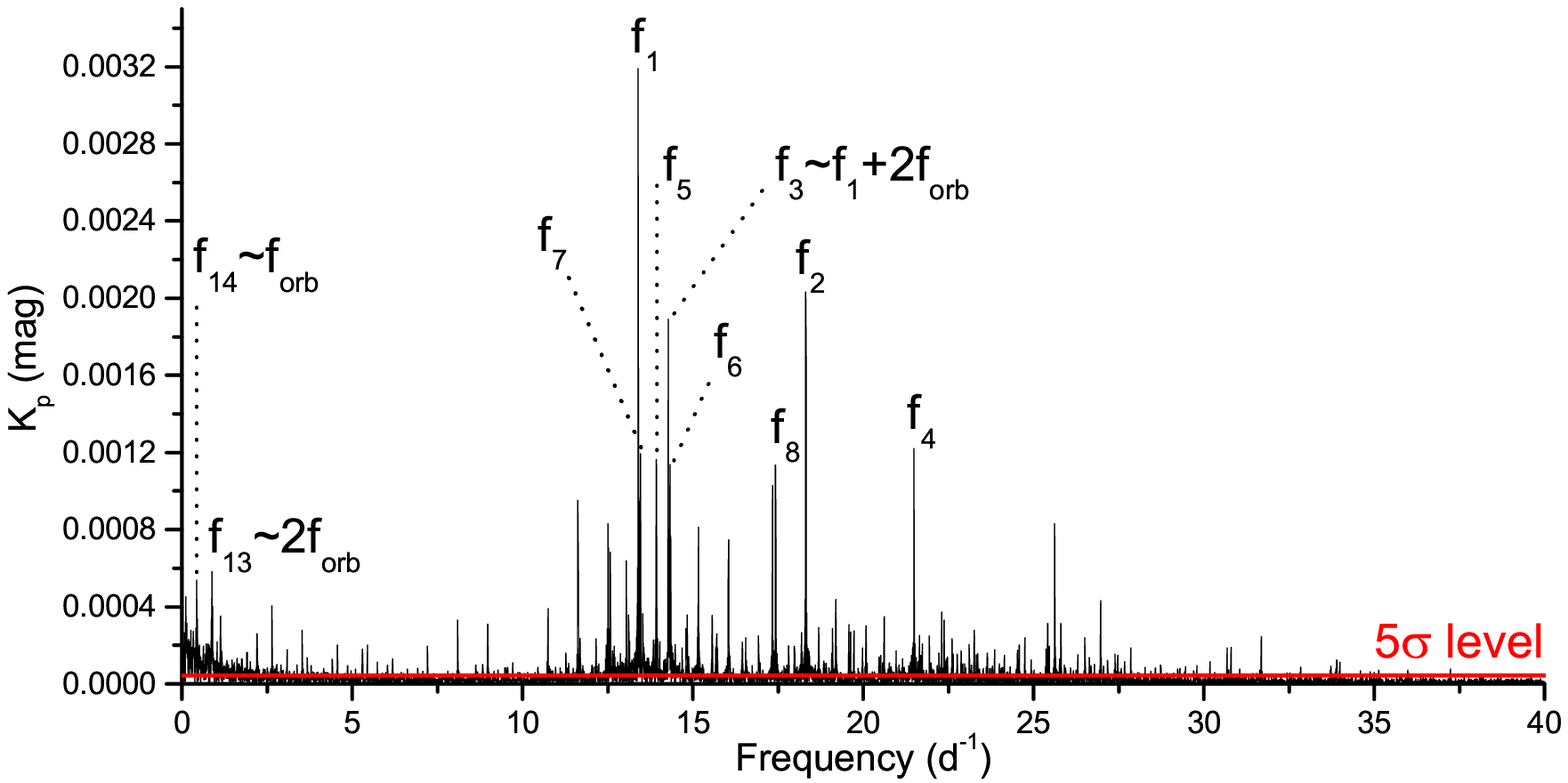}\\
    \includegraphics[width=11cm]{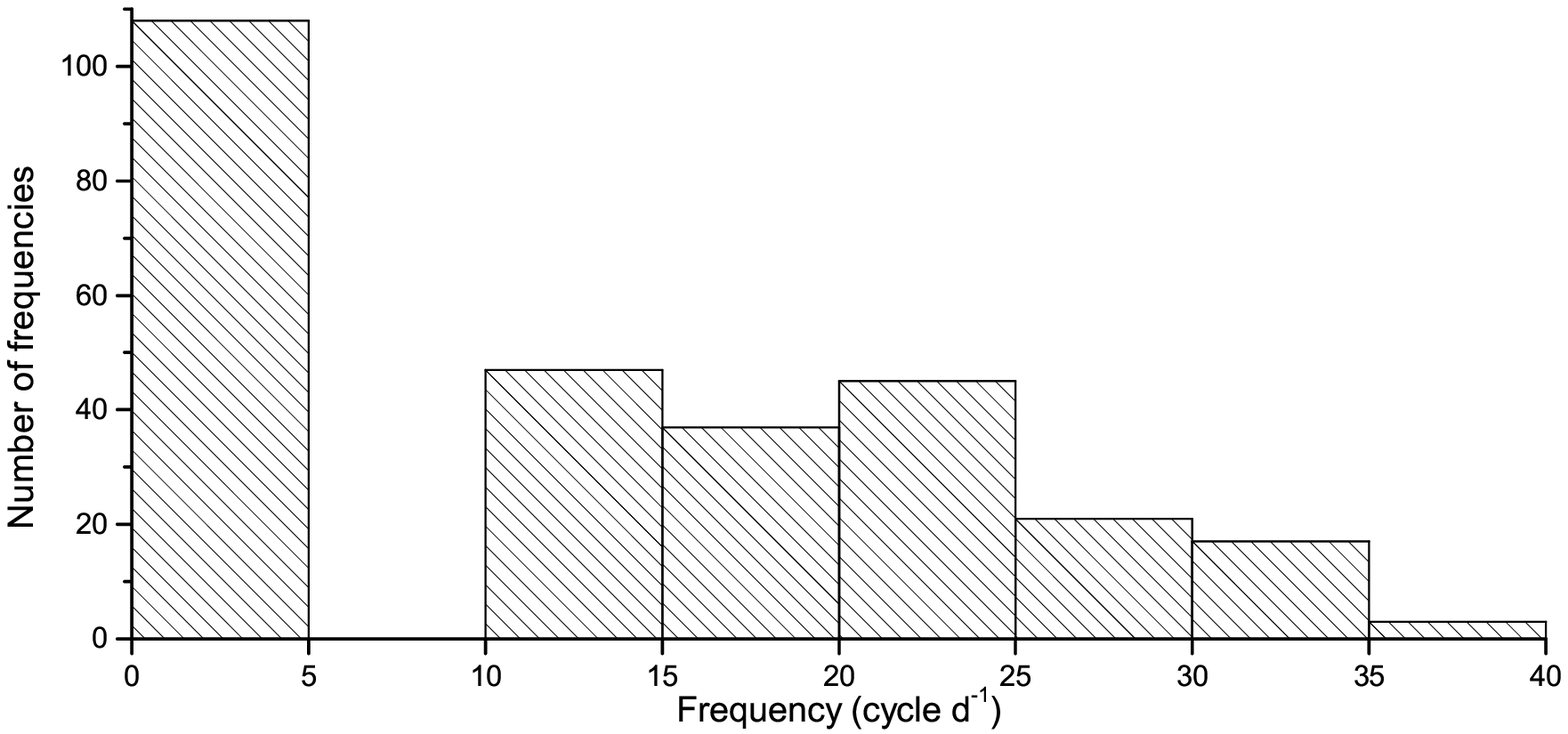}\\
	\end{tabular}
	\FigCap{Periodogram (top) and frequencies distribution (bottom) for the $\delta$~Scuti member of KIC~6629588. The independent frequencies, the strong frequencies that are connected to the $P_{\rm orb}$, and the significance level are also indicated.}
\end{figure}
\begin{figure}
	\centering
		\includegraphics[width=12cm]{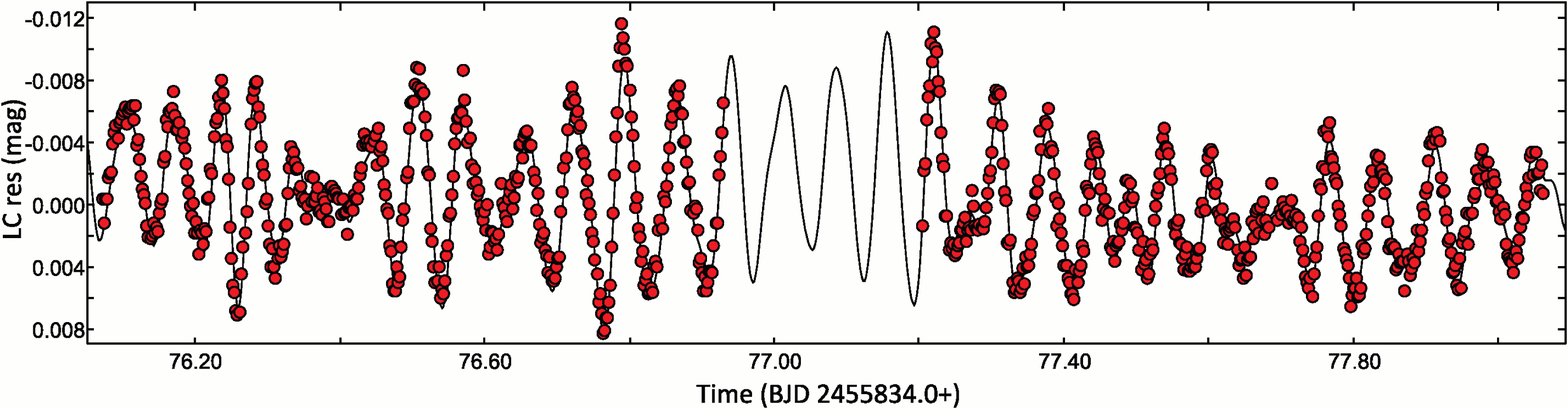}\\
		\FigCap{Representative example of Fourier fitting (solid line) on various data points of KIC~6629588.}
\end{figure}

Table~2 includes the frequency search results regarding the independent frequencies 
Particularly, this table lists: The frequency value $f_{\rm i}$, the amplitude $A$, the phase $\Phi$, the S/N, and the $Q$ values of each detected independent frequency. 
The rest detected (dependent/combination) frequencies are given in Appendix~B (Table~4). Fig.~6 shows the periodogram of the pulsating star and the distribution of its oscillation frequencies, while a representative Fourier fitting on individual LCs residuals is plotted in Fig.~7.


	\MakeTable{l cc c cc}{13cm}{Frequency search results 
for the independent frequencies of the pulsating component of KIC~6629588. The errors are given in parentheses alongside values and correspond to the last digit(s).}
		{\hline																	
		$i$	&	  $f_{\rm i}$	&	$A$	    &	  $\Phi$	&	S/N	  & 	$Q$		\\
		    &	     (d$^{-1}$)	&	(mmag)	&	($^\circ$)	&		  & 	(d) 	\\
		\hline																										
1	&	13.39653(1)	&	3.291(7)	&	39.5(1)	    &	383.6	&	0.034(2)	\\
2	&	18.31601(2)	&	1.882(7)	&	169.7(2)	&	217.1	&	0.025(2)	\\
4	&	21.49330(3)	&	1.217(7)	&	56.2(3)	    &	141.9	&	0.021(1)	\\
5	&	13.93061(4)	&	1.039(7)	&	309.9(4)	&	121.2	&	0.033(2)	\\
6	&	14.32568(4)	&	0.873(7)	&	227.2(4)	&	101.7	&	0.032(2)	\\
7	&	13.46110(5)	&	0.815(7)	&	341.8(5)	&	95.1	&	0.034(2)	\\
8	&	17.33552(4)	&	0.785(7)	&	119.0(4)	&	114.8	&	0.026(2)	\\
		\hline																									
}

As implied in the aforementioned sections, the pulsating star of the system is the primary component. The latter is based on its spectral type as well as to the following results regarding its pulsational behaviour, that are rather typical for a $\delta$~Scuti type pulsator. This star was found to oscillate in seven independent frequencies, while another 271 combination frequencies were also detected. The first ones range between 13.3-21.5~d$^{-1}$, while the majority of the others (163) lie in the regime 10.9-36~d$^{-1}$. The rest 108 frequencies have values less than 5~d$^{-1}$, but they are either combinations of others or are connected to the $f_{\rm orb}$ (\textit{e.g.} $f_{13}$, $f_{14}$, $f_{22}$, $f_{23}$). A check for possible rotational splitting was also performed for the independent frequencies up to the azimuthal order $m=\pm2$ (\textit{i.e.} $m=\pm l$, where $l$ is the spherical degree) assuming a synchronous rotation (\textit{i.e.} $f_{\rm orb}=f_{\rm rot}=0.441604$~d$^{-1}$). The results, which are given in Appendix~C (Table~5), showed that there is no systematic splitting of frequencies due to the rotation . 
The density of this star was calculated as 0.208(3)~$\rho_{\odot}$.

\section{Summary, discussion, and conclusions}

In this study, spectroscopic, photometric, and frequency analyses for the detached eclipsing system KIC~6629588 were presented. The spectral classification of the components of the system played essential role in its LCs modelling, that resulted in the estimation of their absolute parameters and the geometrical configuration of the system. The frequency analysis, in combination with the results from the spectroscopy and photometry, revealed that the primary component is a $\delta$~Scuti type pulsating star and its pulsational properties were accurately determined. This work is a contribution to the topic of Asteroseismolgy of close detached binary systems with $P_{\rm orb}<$13~d, which, so far, includes only 44 known cases.

The components of KIC~6629588 are placed on the mass-radius ($M-R$) and Hertzsprung-Russell ($HR$) evolutionary diagrams in Figs.~8 and 9, respectively, in order to check their evolutionary stages. The primary was classified as an F1V star, it is located inside the main-sequence and it follows very well the theoretical single star evolutionary tracks of Girardi \textit{et al.} (2000) (see Fig.~9) according to its mass and the corresponding error ranges (see Table~1). The main parameters of the used evolutionary tracks, as given in Girardi \textit{et al.} (2000), are: a) the metallicity $Z$=0.019, b) the $He$ abundance $Y=0.273$, c) the $H$ abundance at ZAMS ($X_{c,0}$) ranges between 0.704-0.707, d) the radiative opacities according to the temperatures of the stars, which were taken from Alexander and Ferguson (1994) for $T<10000$~K, and Rogers and Iglesias (1992) and Iglesias and Rogers (1993) for $T>12500$~K, e) the mixing length parameter a$_{\rm MLT}$=1.68, and f) the overshooting parameter $\Lambda_{\rm c}$, which has a zero value for stellar masses less $1~M_{\odot}$, a 0.5 value for stellar masses greater than $1.5~M_{\odot}$, and is between zero and 0.5 for stellar masses between 1-1.5~$M_{\odot}$. This star has similar absolute properties with other $\delta$~Scuti stars in detached binary systems. It is the third less massive and the fourth less luminous star of this sample (\textit{i.e.} detached binary systems with $P_{\rm orb}<$13~d) and it is located very close to the red edge of the classical instability strip. Contrary to this, the secondary is found beyond the terminal age main-sequence (TAMS) indicating that it has been evolved. The fact that it is the less massive but, at the same time, the more evolved component in its system indicates that mass exchange had been occurred in the past with a direction from the present secondary to the primary. The most probable evolutionary scenario is that the mass donor (current secondary) was initially the more massive component, evolved faster, reached its critical Roche radius, hence, the mass transfer began. Finally, after the mass ratio of the system inverted, the mass exchange stopped. Moreover, the current secondary component was found to be magnetically active (\textit{i.e.} a cool spot on its surface was needed for the LCs modelling) during the selected three-month period of data coverage.

\begin{figure}
\begin{center}
\includegraphics[width=11cm]{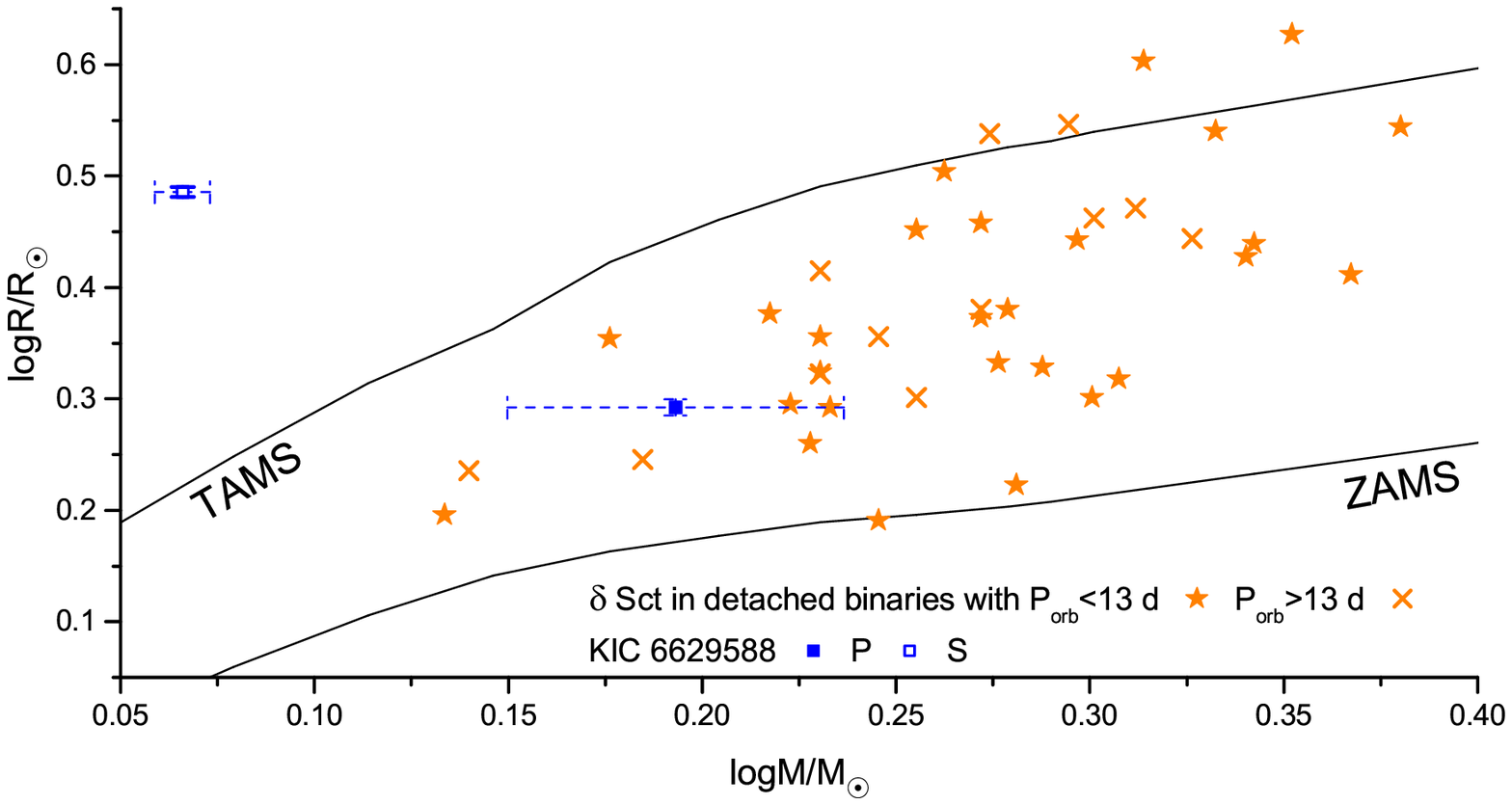}
\end{center}
\FigCap{Location of the primary (filled square) and secondary (empty square) components of KIC~6629588 within the mass-radius diagram. The stars and the `x' symbols denote the $\delta$~Scuti components of other detached systems (taken from Liakos and Niarchos 2017 and Liakos 2020) with $P_{\rm orb}$ shorter and longer than 13~d, respectively. The black solid lines represent the main-sequence edges.}
\end{figure}
\begin{figure}[h]
\begin{center}
\includegraphics[width=11cm]{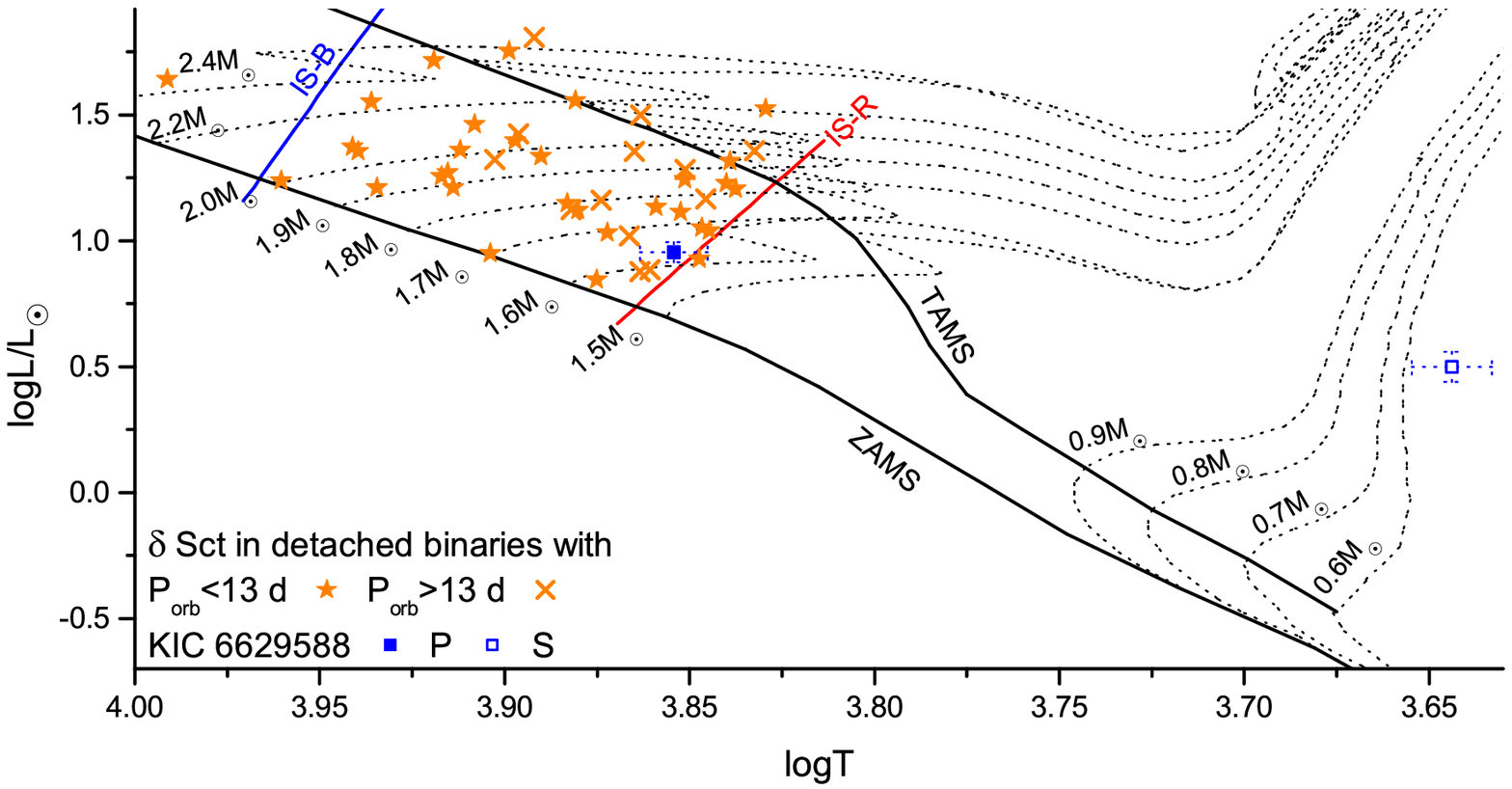}
\end{center}
\FigCap{Locations of the components of KIC~6629588 within the $HR$ diagram. Symbols and black solid lines have the same meaning as in Fig.~8. Dotted lines denote the evolutionary tracks for single stars with masses between 0.6-0.9~$M_{\odot}$ and 1.5-2.4~$M_{\odot}$ (taken from Girardi \textit{et al.} 2000) and the coloured solid lines (B=Blue, R=Red) the boundaries of the instability strip (IS; taken from Murphy \textit{et al.} 2019).}
\end{figure}

The primary of the system oscillates in seven independent frequencies, with the dominant at $\sim$13.4~d$^{-1}$. 
In order to compare its pulsational properties with other similar cases, it is placed on the $P_{\rm pul}-P_{\rm orb}$ and $\log g - P_{\rm pul}$ diagrams (Figs~10 and 11, respectively), that contain the total sample and the relations of Liakos (2020) for $\delta$~Scuti stars in detached binaries with $P_{\rm orb}<13$~d. These plots show that the $\delta$~Scuti member of KIC~6629588 follows very well both the distributions of the sample stars as well as the empirical relations.

\begin{figure}[t]
\begin{center}
\includegraphics[width=10.5cm]{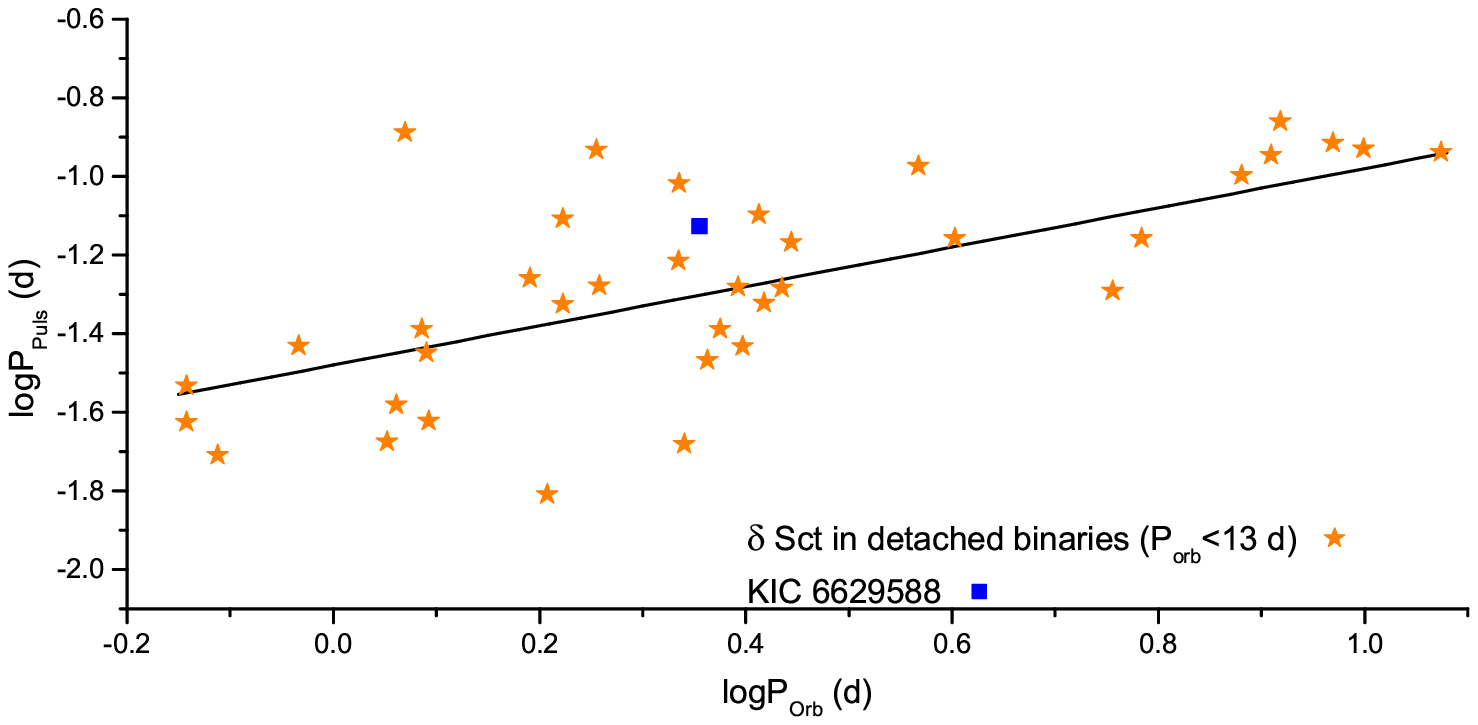}
\end{center}
	\FigCap{Location of the oscillating component of KIC~6629588 among other $\delta$~Scuti stars-members of detached systems with $P_{\rm orb}<13$~d within the $P_{\rm orb}-P_{\rm pul}$ diagram. Symbols have the same meaning as in Figs~8 and 9, while the solid line denotes the empirical linear fit of Liakos (2020).}
\begin{center}
\includegraphics[width=10.5cm]{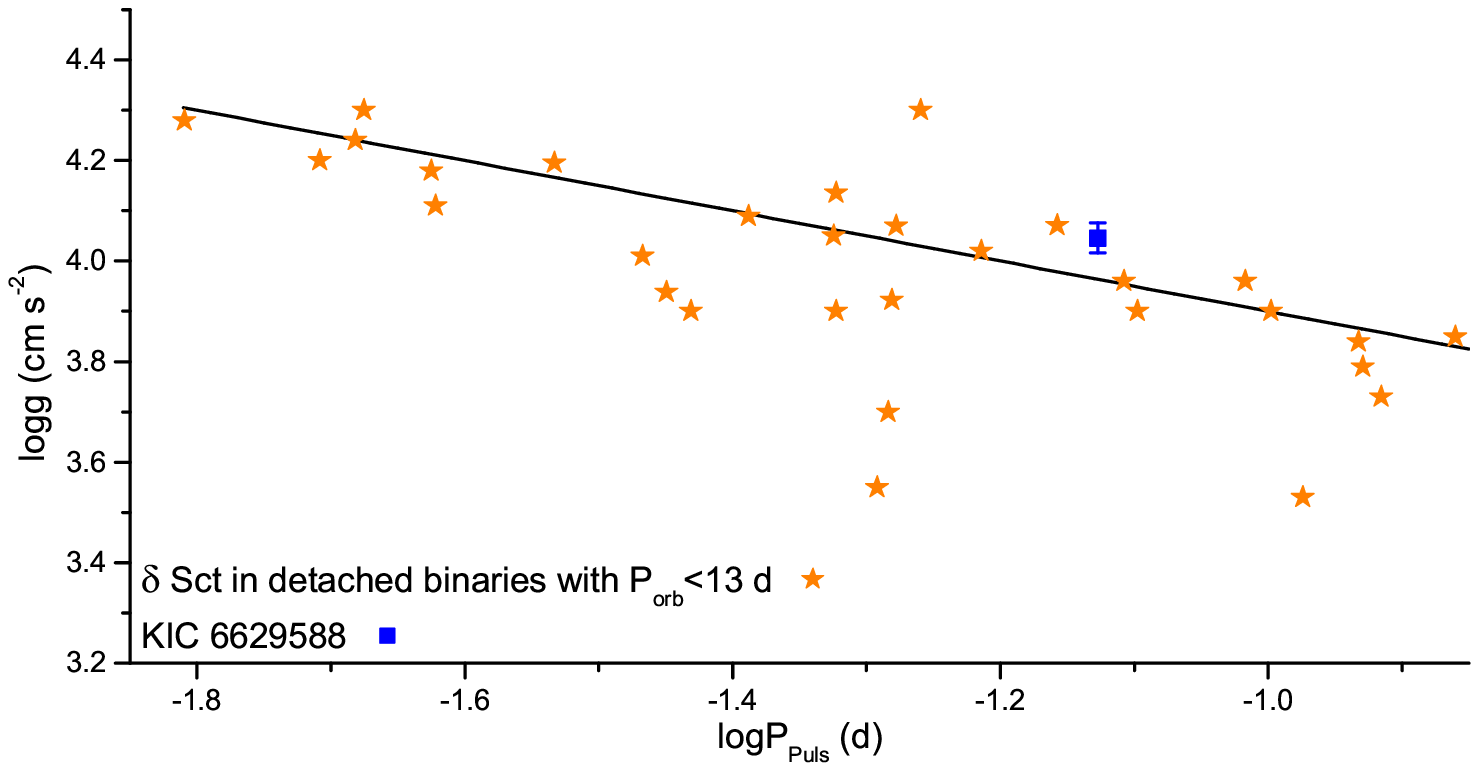}
\end{center}
	\FigCap{Location of the $\delta$~Scuti star of KIC~6629588 within the $\log g-P_{\rm pul}$ diagram. Symbols and lines have the same meaning as in Fig.~10.}
\end{figure}

Based on the pulsation period-luminosity relation for $\delta$~Scuti stars of Ziaali \textit{et al.} (2019):
\begin{equation}
M_{\rm V} = -2.94(6) \log P_{\rm pul}-1.34(6),
\end{equation}
it is feasible to calculate the absolute magnitude ($M_{\rm V}$) of the oscillating component of KIC~6629588 using its dominant frequency. Subsequently, using the apparent magnitude $m_{\rm V}$=13.88~mag (NOMAD-1 catalog; Zacharias \textit{et al.} 2004) and the extinction $A_{\rm V}=$~0.644~mag (Berger \textit{et al.} 2018) of the system, and the distance modulus, the distance can be determined as 1789$^{+102}_{-108}$~pc. This value is in good agreement with that of Anders \textit{et al.} (2019; 1884~pc) and in less good agreement with that of Berger \textit{et al.} (2018; 2018$\pm91$~pc) and Bailer-Jones \textit{et al.} (2018; 2017$^{+93}_{-85}$~pc). The distance values in these studies were calculated combining the parallax value ($\pi$) of the GAIA DR2 catalog (Gaia Collaboration \textit{et al.} 2018) and other photometric measurements. However, using directly the parallax value as given in GAIA DR2 ($\pi_{\rm DR2}=0.468\pm0.022$~mas), the distance is derived as $2138\pm99$~pc, while using the respective value of GAIA EDR3 (Gaia Collaboration 2020), \textit{i.e.} $\pi_{\rm EDR3}=0.509\pm0.13$~mas, the distance yields $1965\pm50$~pc.

It should be noticed that the $M_{\rm V}$ as derived from Eq.~1, hence the distance, is reliable only if the dominant frequency is a radial mode. Therefore, in order to check also if $f_1$ is indeed a radial mode, the $M_{\rm V}$ is also calculated using the $M_{\rm bol, 1}$ from Table~1 and the bolometric correction ($BC$) of Flower (1996), \textit{i.e.} $BC$=0.034~mag for $T_{\rm eff}=$7150~K. The latter calculation results in $M_{\rm V}=2.33\pm0.05$~mag, while the respective value using Eq.~1 is $1.97\pm0.13$~mag. The comparison between these values, although they do not have large difference, indicates that probably $f_1$ is not a radial mode.

Considering potential future studies for this system, radial velocity measurements are welcome to validate the present results regarding the EB model. However, the latter can be considered as a relatively difficult task, because for their acquisition, a telescope with diameter 4+~m equipped with a high-resolution spectrograph is needed due to the system's relatively faint magnitude ($\sim14$~mag). In any case, it is anticipated that the radial velocity curves cannot significantly change the present pulsational properties. It is encouraged and strongly recommenced for other researchers to contribute with asteroseismic studies of other similar systems, in order to enlarge the current small sample. Future works on this kind of systems will offer the opportunity to enrich our knowledge about the influence of binarity on pulsations, and will help to come closer to the answer for the critical question about the evolution of pulsations in the presence of a close component.

\Acknow{The author acknowledges financial support by the European Space Agency (ESA) under the Near Earth object Lunar Impacts and Optical TrAnsients (NELIOTA) programme, contract no. 4000112943, and wishes to thank Mrs Maria Pizga for proofreading the text and the anonymous reviewer for the valuable comments. The `Aristarchos' telescope is operated on Helmos Observatory by the Institute for Astronomy, Astrophysics, Space Applications and Remote Sensing of the National Observatory of Athens. This research has made use of NASA's Astrophysics Data System Bibliographic Services, the SIMBAD, the Mikulski Archive for Space Telescopes (MAST), and the $Kepler$ Eclipsing Binary Catalog data bases. This work has made use of data from the European Space Agency (ESA) mission Gaia (https://www.cosmos.esa.int/gaia), processed by the Gaia Data Processing and Analysis Consortium (DPAC). Funding for the DPAC has been provided by national institutions, in particular the institutions participating in the Gaia Multilateral Agreement.}

\begin{appendix}

\section{Spot migration}
This appendix includes information for the time-depended variations of the spot, which was detected on the secondary component of the system (see also Section~3). The average BJD values of the points included in the models, from which the respective parameters (latitude $Co-lat$, longitude $long$, radius, and temperature factor $Tf$) were calculated, were set as corresponding timings for each group of cycles in Table~3. The upper part of Fig.~12 shows the changes of the parameters of the spot over time, while the lower parts show the spot on the secondary's surface during two different days of observations.

\MakeTable{c cc cc c cc cc}{14cm}{Spot parameters for KIC~6629588.}
{\hline	
Time$^{(1)}$	&	Co-lat	&	Long.	&	Radius	&	Tf ($\frac{T_{\rm spot}}{T_{\rm eff}}$)	&	Time$^{(1)}$&	Co-lat	&	Long.	&	Radius	&	Tf ($\frac{T_{\rm spot}}{T_{\rm eff}}$)	\\
(days)	&	($^\circ$)	&	($^\circ$)	&	($^\circ$)	&		&	(days)	&	($^\circ$)	&	($^\circ$)	&	($^\circ$)	&		\\
\hline
0.60	&	23(1)	&	129(1)	&	24(1)	&	0.87(1)	&	48.15	&	25(1)	&	131(1)	&	27(1)	&	0.88(1)	\\
7.39	&	23(1)	&	131(2)	&	24(1)	&	0.87(1)	&	54.89	&	26(1)	&	134(0)	&	27(1)	&	0.88(1)	\\
14.18	&	23(1)	&	127(1)	&	25(1)	&	0.87(1)	&	62.04	&	26(1)	&	135(1)	&	26(1)	&	0.87(1)	\\
20.99	&	27(1)	&	130(2)	&	26(1)	&	0.90(1)	&	70.53	&	26(1)	&	128(1)	&	27(1)	&	0.87(1)	\\
27.61	&	27(1)	&	130(1)	&	27(1)	&	0.90(1)	&	77.59	&	27(1)	&	123(1)	&	26(1)	&	0.85(1)	\\
34.56	&	27(1)	&	136(1)	&	28(1)	&	0.90(1)	&	84.36	&	26(1)	&	122(1)	&	26(1)	&	0.85(1)	\\
41.35	&	26(1)	&	127(1)	&	28(1)	&	0.90(1)	&	91.05	&	25(1)	&	125(1)	&	27(1)	&	0.84(1)	\\
\hline
\multicolumn{10}{p{10cm}}{$^{(1)}$The starting BJD is: 2455837.0+}
}



\begin{figure}[h]
\begin{center}
\begin{tabular}{cr}
		\multicolumn{2}{c}{\includegraphics[width=7cm]{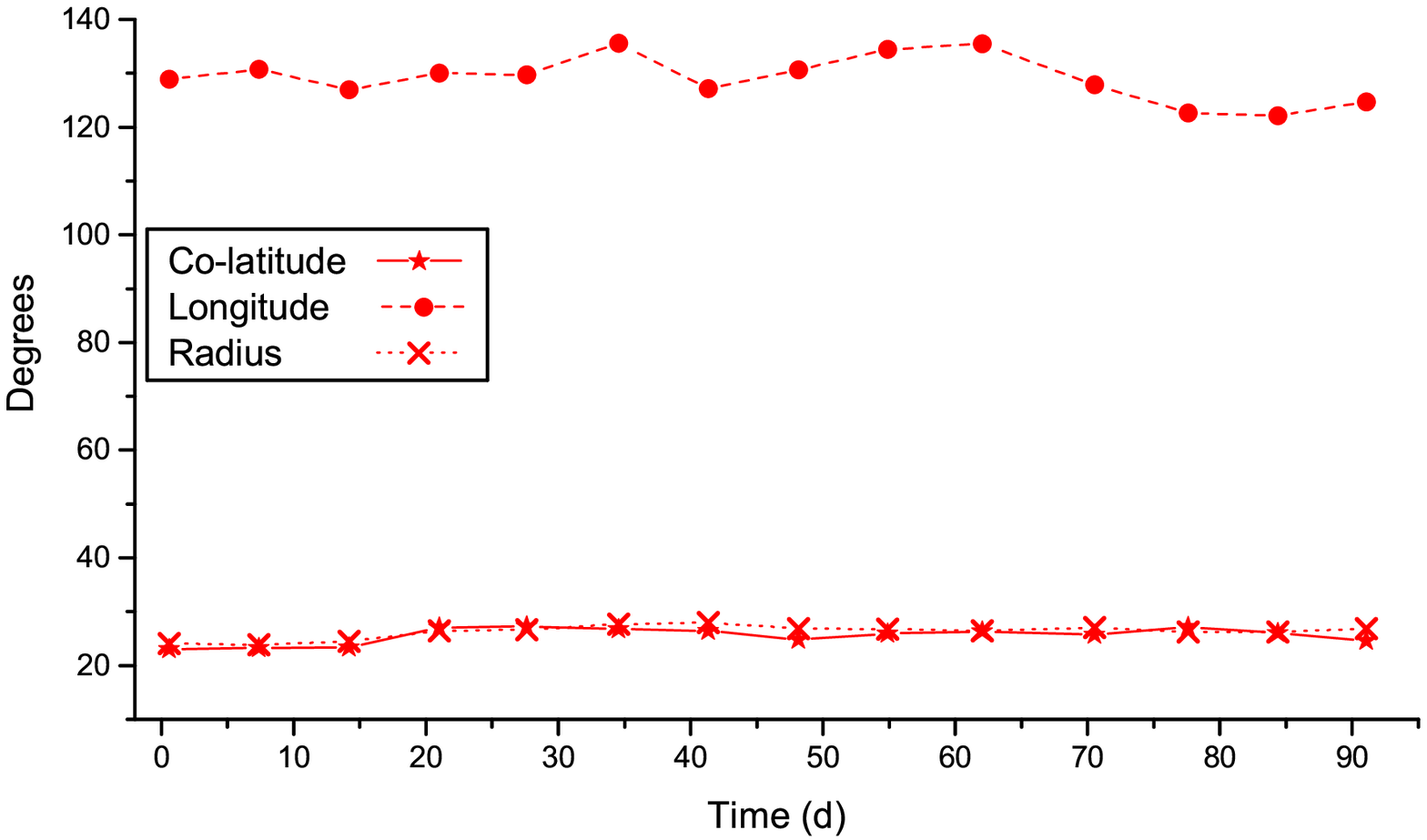}}\\
		\includegraphics[width=2.5cm]{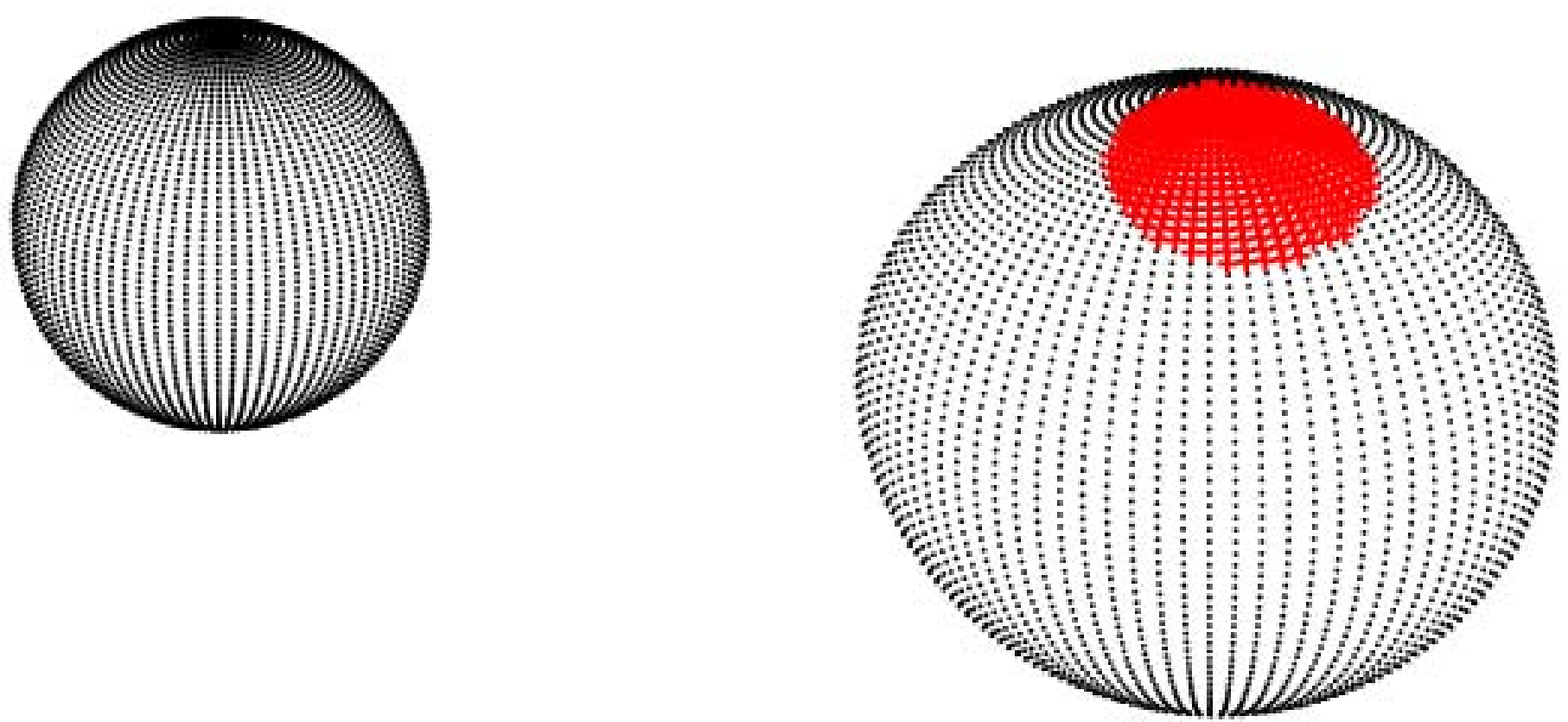}&\includegraphics[width=2.5cm]{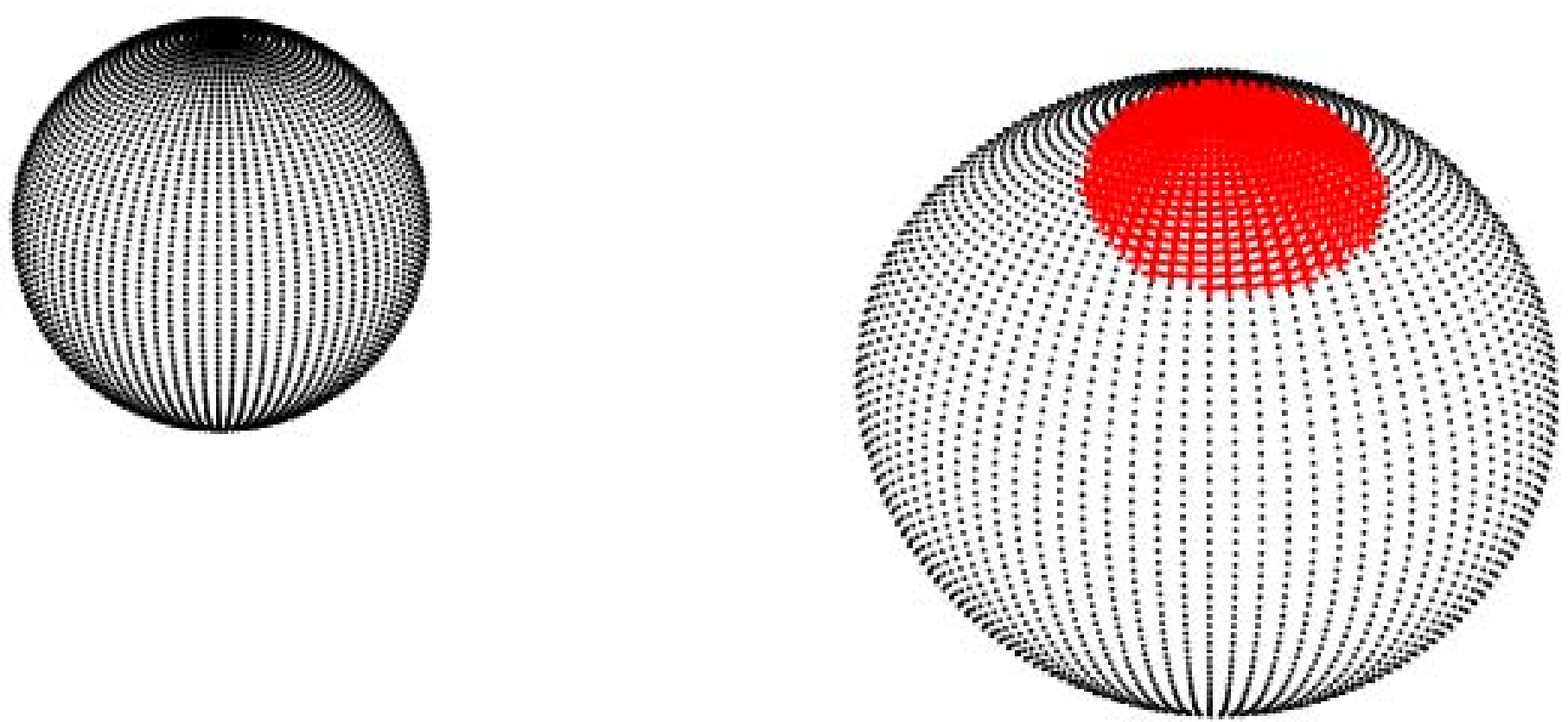}\\
	\end{tabular}
\end{center}
	\FigCap{Upper panel: Spot migration diagram for KIC~6629588. Lower panels: The location of the cool spot (red area) on the surface of the secondary component during the first (left) and the last days (right) of observations, when the system is at orbital phase 0.18.}
\end{figure}

\section{Combination Frequencies}

Table~4 contains the values of the depended frequencies $f_{\rm i}$ (where $i$ is an increasing number), semi-amplitudes $A$, phases $\Phi$, and S/N. Moreover, in the last column of this table, the most possible combination for each frequency is also given. The combinations were calculated only for the first 255 frequencies because this is the maximum number of frequencies that the software can detect during one run (\textit{i.e.} Fourier model). However, it should be noted that, in order to continue the search, the residuals from these Fourier models were given as new data sets to the same software, but no combinations could be calculated using the first 255 frequencies. It should be also noticed that frequencies with similar values (inside the frequency resolution range) to others with higher amplitudes may be possible leftovers of them. However, given that there is not any accurate method to distinguish this possibility from a simple coincidence of similarity of two true close frequencies, these frequencies are also listed in the following table.

\begin{center}
\begin{longtable}{cc cc cl}
\caption{Combination frequencies of KIC~6629588.}\\
\hline
$i$~~~	&~~~~~	$f_{\rm	i}$~~~~~~~~~	&	~~~~~~$A$~~	     &	~~~~~~~~~~$\Phi$~~~~~~~    &	~~~S/N~~	&	Combination	~~~~~~~			\\
~~~~	&~~~~~(d$^{-1}$)~~~~~~~~~	&	~~~~(mmag)~~	&	~~~~~~~~~~($^\circ$)~~~~~~~ 	&	~~~~~~~~    &						\\
\hline
\endfirsthead
\multicolumn{6}{c}%
{\tablename\ \thetable\ -- \textit{Continued from previous page}} \\
\hline
$i$	&	$f_{\rm	i}$	&	$A$	&	$\Phi$	&	S/N	&	Combination				\\
	&	(d$^{-1}$)	&	(mmag)	&	($^\circ$)	&	&						\\
\hline
\endhead
\hline \\
\endfoot
\hline
\endlastfoot
3	&	14.27990(2)	&	1.869(7)	&	12.8(2)	&	220.2	&	$f_1$+$2f_{\rm orb}$	\\
9	&	25.62660(5)	&	0.816(7)	&	9.0(5)	&	95.2	&	3$f_6-f_8$	\\
10	&	13.92145(5)	&	0.721(7)	&	165.9(5)	&	84.0	&	$\sim f_5$	\\
11	&	13.44204(5)	&	0.721(7)	&	78.1(5)	&	84.1	&	$f_1$+$f_6-f_3$	\\
12	&	14.34577(6)	&	0.632(7)	&	142.2(6)	&	73.6	&	$f_3$+$f_7-f_1$	\\
13	&	0.88313(4)	&	0.993(7)	&	221.2(4)	&	115.8	&	$2f_{\rm orb}$	\\
14	&	0.44131(8)	&	0.497(7)	&	45.1(8)	&	58.0	&	$f_{\rm orb}$	\\
15	&	17.43284(7)	&	0.543(7)	&	11.4(7)	&	63.4	&	$f_2-f_{13}$	\\
16	&	0.11071(7)	&	0.510(7)	&	7.8(7)	&	59.5	&	$f_{\rm orb}-f_{8}$	\\
17	&	13.94051(8)	&	0.458(7)	&	133.4(8)	&	53.4	&	$\sim f_5$	\\
18	&	26.97008(9)	&	0.432(7)	&	281.7(9)	&	50.4	&	$f_1$+$f_{16}$+$f_7$	\\
19	&	12.58054(7)	&	0.519(7)	&	85.3(7)	&	60.5	&	$f_{7}-f_{13}$	\\
20	&	0.12359(8)	&	0.445(7)	&	100.2(9)	&	51.8	&	$\sim f_{16}$	\\
21	&	22.30007(11)	&	0.352(7)	&	236.8(1.1)	&	41.0	&	$2 f_{2}-f_6$	\\
22	&	1.76522(6)	&	0.632(7)	&	231.0(6)	&	73.7	&	$4f_{\rm orb}$	\\
23	&	0.89754(11)	&	0.357(7)	&	29.8(1.1)	&	41.6	&	$2f_{\rm orb}$	\\
24	&	1.13905(12)	&	0.322(7)	&	9.6(1.2)	&	37.6	&	$f_{13}$+$2f_{20}$	\\
25	&	14.83753(11)	&	0.332(7)	&	292.3(1.1)	&	38.7	&	$4f_{\rm orb}$+$f_{5}$	\\
26	&	14.31435(11)	&	0.328(7)	&	162.3(1.2)	&	38.3	&	$\sim f_6$	\\
27	&	25.41290(12)	&	0.309(7)	&	97.6(1.2)	&	36.1	&	$f_{\rm orb}$+$f_{21}-f_6$	\\
28	&	0.14779(12)	&	0.318(7)	&	281.5(1.2)	&	37.1	&	$f_{11}$+$f_{16}-f_1$	\\
29	&	25.81043(12)	&	0.314(7)	&	160.1(1.2)	&	36.6	&	$f_{27}$+$f_{6}-f_5$	\\
30	&	13.04554(11)	&	0.345(7)	&	137.4(1.1)	&	40.3	&	$f_{5}-f_{13}$	\\
31	&	0.46036(13)	&	0.296(7)	&	172.3(1.3)	&	34.5	&	$f_{5}-f_{7}$	\\
32	&	13.12072(13)	&	0.296(7)	&	351.4(1.3)	&	34.6	&	$f_{13}$+$f_{9}-f_1$	\\
33	&	0.26777(13)	&	0.288(7)	&	75.0(1.3)	&	33.5	&	$f_{1}-f_{32}$	\\
34	&	0.13491(13)	&	0.288(7)	&	165.3(1.3)	&	33.6	&	$\sim f_{20}$	\\
35	&	18.19444(14)	&	0.274(7)	&	165.6(1.4)	&	31.9	&	$f_{2}-f_{16}$	\\
36	&	16.05898(14)	&	0.270(7)	&	208.1(1.4)	&	31.5	&	$f_{22}$+$f_{3}$	\\
37	&	21.65125(14)	&	0.263(7)	&	146.1(1.4)	&	30.7	&	$f_{28}$+$f_{4}$	\\
38	&	31.68025(15)	&	0.258(7)	&	122.3(1.5)	&	30.1	&	$f_{12}$+$f_{8}$	\\
39	&	19.62598(15)	&	0.256(7)	&	182.4(1.5)	&	29.9	&	$f_{13}$+$2f_{\rm orb}$+$f_2$	\\
40	&	0.08085(15)	&	0.249(7)	&	18.9(1.5)	&	29.0	&	$f_{12}-f_{3}$	\\
41	&	14.02547(15)	&	0.248(7)	&	350.1(1.5)	&	28.9	&	$f_{3}-f_{33}$	\\
42	&	0.50979(15)	&	0.245(7)	&	97.3(1.6)	&	28.6	&	$f_{10}-f_{1}$	\\
43	&	0.25129(15)	&	0.246(7)	&	97.1(1.5)	&	28.7	&	$2f_{20}$	\\
44	&	0.23687(14)	&	0.261(7)	&	42.7(1.5)	&	30.4	&	$\sim f_{43}$	\\
45	&	1.32237(16)	&	0.237(7)	&	106.4(1.6)	&	27.6	&	$3f_{\rm orb}$	\\
46	&	24.52926(16)	&	0.234(7)	&	129.1(1.6)	&	27.2	&	$f_{27}-f_{13}$	\\
47	&	13.53164(16)	&	0.231(7)	&	47.2(1.6)	&	26.9	&	$f_{1}$+$f_{20}$	\\
48	&	22.61316(16)	&	0.228(7)	&	55.7(1.7)	&	26.6	&	$2f_{2}-f_{41}$	\\
49	&	0.87386(17)	&	0.226(7)	&	100.1(1.7)	&	26.3	&	$\sim f_{13}$	\\
50	&	18.21864(17)	&	0.225(7)	&	99.3(1.7)	&	26.3	&	$f_{13}$+$f_{8}$	\\
51	&	21.94785(17)	&	0.224(7)	&	79.5(1.7)	&	26.1	&	$2f_{\rm orb}$+$f_{4}$	\\
52	&	24.57766(18)	&	0.210(7)	&	297.8(1.8)	&	24.5	&	$f_{11}$+$3f_{\rm orb}-f_1$	\\
53	&	0.33935(18)	&	0.208(7)	&	258.4(1.8)	&	24.2	&	$f_{3}-f_{5}$	\\
54	&	17.98177(18)	&	0.206(7)	&	39.2(1.8)	&	24.0	&	$f_{2}-f_{53}$	\\
55	&	0.21061(19)	&	0.201(7)	&	281.5(1.9)	&	23.5	&	$2f_{16}$	\\
56	&	15.67998(19)	&	0.201(7)	&	272.9(1.9)	&	23.5	&	$f_{10}$+$f_{22}$	\\
57	&	0.36355(19)	&	0.201(7)	&	64.9(1.9)	&	23.4	&	$f_{3}-f_{5}$	\\
58	&	14.48274(19)	&	0.195(7)	&	192.9(1.9)	&	22.8	&	$f_{3}$+$f_{55}$	\\
59	&	0.42019(20)	&	0.188(7)	&	228.8(2.0)	&	21.9	&	$2f_{55}$	\\
60	&	0.17045(19)	&	0.194(7)	&	59.5(2.0)	&	22.6	&	$2f_{40}$	\\
61	&	0.91299(21)	&	0.182(7)	&	123.2(2.1)	&	21.2	&	$2f_{31}$	\\
62	&	25.37222(21)	&	0.181(7)	&	81.6(2.1)	&	21.1	&	$f_{9}-f_{33}$	\\
63	&	30.67869(21)	&	0.182(7)	&	317.3(2.1)	&	21.2	&	$f_{1}$+$f_{\rm orb}-f_{28}$	\\
64	&	12.46262(21)	&	0.181(7)	&	248.1(2.1)	&	21.1	&	$f_{19}-f_{16}$	\\
65	&	0.10402(22)	&	0.173(7)	&	148.8(2.2)	&	20.2	&	$\sim f_{16}$	\\
66	&	22.62037(22)	&	0.170(7)	&	44.9(2.2)	&	19.8	&	$\sim f_{48}$	\\
67	&	13.83855(22)	&	0.168(7)	&	343.0(2.3)	&	19.6	&	$f_{1}$+$2f_{\rm orb}$	\\
68	&	22.87938(24)	&	0.158(7)	&	0.5(2.4)	&	18.4	&	$f_{33}$+$f_{48}$	\\
69	&	12.87355(24)	&	0.156(7)	&	129.9(2.4)	&	18.2	&	$f_{1}-f_{42}$	\\
70	&	0.49331(24)	&	0.154(7)	&	297.6(2.5)	&	18.0	&	$2f_{43}$	\\
71	&	0.52833(23)	&	0.162(7)	&	40.1(2.4)	&	18.8	&	$2f_{33}$	\\
72	&	0.19465(25)	&	0.154(7)	&	213.3(2.5)	&	17.9	&	$2f_{65}$	\\
73	&	0.07209(25)	&	0.149(7)	&	119.8(2.6)	&	17.4	&	$\sim f_{40}$	\\
74	&	0.04274(26)	&	0.146(7)	&	65.7(2.6)	&	17.0	&	$f_{11}-f_{1}$	\\
75	&	0.83833(23)	&	0.167(7)	&	141.5(2.3)	&	19.5	&	$2f_{59}$	\\
76	&	21.45093(26)	&	0.143(7)	&	103.2(2.7)	&	16.7	&	$f_{4}-f_{74}$	\\
77	&	21.68317(26)	&	0.143(7)	&	30.6(2.7)	&	16.7	&	$f_{4}$+$f_{72}$	\\
78	&	23.36343(27)	&	0.141(7)	&	161.5(2.7)	&	16.4	&	$f_{68}$+$f_{70}$	\\
79	&	22.47927(27)	&	0.142(7)	&	129.7(2.7)	&	16.5	&	$f_{78}$+$f_{13}$	\\
80	&	27.47060(27)	&	0.138(7)	&	144.7(2.7)	&	16.1	&	$f_{47}$+$f_{5}$	\\
81	&	13.91321(28)	&	0.135(7)	&	153.1(2.8)	&	15.8	&	$\sim f_{10}$	\\
82	&	0.43152(28)	&	0.134(7)	&	285.7(2.8)	&	15.7	&	$\sim 2f_{\rm orb}$	\\
83	&	0.40680(26)	&	0.147(7)	&	192.2(2.6)	&	17.2	&	$\sim f_{59}$	\\
84	&	1.17767(28)	&	0.133(7)	&	283.8(2.9)	&	15.5	&	$4f_{\rm orb}$+$f_{33}$	\\
85	&	16.44982(28)	&	0.133(7)	&	192.3(2.9)	&	15.5	&	$f_{8}-f_{13}$	\\
86	&	21.15433(29)	&	0.129(7)	&	283.4(3.0)	&	15	&	$f_{4}-f_{53}$	\\
87	&	0.85686(30)	&	0.128(7)	&	118.9(3.0)	&	14.9	&	$2f_{82}$	\\
88	&	13.96110(30)	&	0.126(7)	&	106.3(3.0)	&	14.7	&	$f_{25}-f_{13}$	\\
89	&	23.32326(30)	&	0.125(7)	&	239.8(3.0)	&	14.6	&	$2f_{\rm orb}$+$f_{68}$	\\
90	&	17.30359(30)	&	0.124(7)	&	278.8(3.1)	&	14.5	&	$f_{35}-f_{13}$	\\
91	&	0.32750(30)	&	0.124(7)	&	168.9(3.1)	&	14.4	&	$\sim f_{53}$	\\
92	&	0.15706(30)	&	0.124(7)	&	50.5(3.1)	&	14.4	&	$\sim f_{28}$	\\
93	&	33.90326(31)	&	0.121(7)	&	306.4(3.1)	&	14.1	&	$f_{3}$+$f_{39}$	\\
94	&	0.28476(32)	&	0.117(7)	&	79.1(3.2)	&	13.7	&	$2f_{28}$	\\
95	&	21.73312(32)	&	0.117(7)	&	275.1(3.2)	&	13.6	&	$f_{4}$+$f_{44}$	\\
96	&	3.53302(32)	&	0.116(7)	&	349.8(3.3)	&	13.6	&	$8f_{\rm orb}$	\\
97	&	11.36116(33)	&	0.115(7)	&	220.0(3.3)	&	13.5	&	$f_{9}-f_{3}$	\\
98	&	13.54091(33)	&	0.114(7)	&	356.7(3.3)	&	13.3	&	$\sim f_{47}$	\\
99	&	22.41748(33)	&	0.114(7)	&	280.9(3.3)	&	13.2	&	$f_{4}$+$f_{61}$	\\
100	&	33.98977(33)	&	0.113(7)	&	13.8(3.4)	&	13.1	&	$f_{2}$+$f_{56}$	\\
101	&	17.33449(34)	&	0.112(7)	&	154.8(3.4)	&	13	&	$\sim f_{8}$	\\
102	&	0.47169(34)	&	0.111(7)	&	137.9(3.4)	&	12.9	&	$\sim f_{31}$	\\
103	&	0.00927(34)	&	0.112(7)	&	71.9(3.4)	&	13.1	&	$f_{5}-f_{10}$	\\
104	&	21.98853(34)	&	0.110(7)	&	253.4(3.4)	&	12.9	&	$f_{4}$+$f_{42}$	\\
105	&	13.05429(35)	&	0.109(7)	&	54.8(3.5)	&	12.7	&	$\sim f_{30}$	\\
106	&	1.05100(35)	&	0.109(7)	&	3.8(3.5)	&	12.7	&	$2f_{71}$	\\
107	&	33.73384(36)	&	0.106(7)	&	283.7(3.6)	&	12.3	&	$f_{19}$+$f_{86}$	\\
108	&	13.40703(36)	&	0.104(7)	&	232.9(3.6)	&	12.1	&	$\sim f_{1}$	\\
109	&	1.68952(37)	&	0.103(7)	&	346.4(3.7)	&	12	&	$2f_{75}$	\\
110	&	0.34913(37)	&	0.103(7)	&	16.4(3.7)	&	12	&	$\sim f_{53}$	\\
111	&	27.67658(37)	&	0.102(7)	&	160.3(3.7)	&	11.9	&	$f_{1}$+$f_{3}$	\\
112	&	13.13668(37)	&	0.102(7)	&	284.1(3.7)	&	11.9	&	$f_{1}-f_{33}$	\\
113	&	11.69896(37)	&	0.101(7)	&	168.6(3.7)	&	11.8	&	$f_{9}-f_{5}$	\\
114	&	0.44234(38)	&	0.100(7)	&	44.8(3.8)	&	11.7	&	$\sim f_{\rm orb}$	\\
115	&	0.03090(38)	&	0.100(7)	&	295.0(3.8)	&	11.6	&	$\sim f_{74}$	\\
116	&	30.79713(39)	&	0.098(7)	&	329.9(3.9)	&	11.4	&	$f_{7}$+$f_{8}$	\\
117	&	21.14557(39)	&	0.098(7)	&	349.0(3.9)	&	11.4	&	$\sim f_{86}$	\\
118	&	22.33354(39)	&	0.097(7)	&	92.1(3.9)	&	11.3	&	$f_{4}$+$f_{75}$	\\
119	&	1.12927(39)	&	0.097(7)	&	217.5(3.9)	&	11.3	&	$\sim f_{24}$	\\
120	&	18.20783(39)	&	0.096(7)	&	213.9(4.0)	&	11.2	&	$\sim f_{35}$	\\
121	&	17.96632(40)	&	0.094(7)	&	39.9(4.1)	&	10.9	&	$f_{2}-f_{53}$	\\
122	&	0.56747(41)	&	0.092(7)	&	245.3(4.1)	&	10.8	&	$2f_{94}$	\\
123	&	0.05149(41)	&	0.092(7)	&	319.8(4.1)	&	10.7	&	$\sim f_{74}$	\\
124	&	25.46079(41)	&	0.092(7)	&	88.2(4.1)	&	10.7	&	$f_{13}$+$f_{52}$	\\
125	&	17.76910(42)	&	0.090(7)	&	245.4(4.2)	&	10.5	&	$2f_{\rm orb}$+$f_{8}$	\\
126	&	22.18009(42)	&	0.090(7)	&	338.7(4.2)	&	10.4	&	$f_{21}-f_{16}$	\\
127	&	1.30950(42)	&	0.089(7)	&	226.1(4.3)	&	10.4	&	$\sim f_{45}$	\\
128	&	13.03627(42)	&	0.089(7)	&	241.7(4.3)	&	10.3	&	$\sim f_{30}$	\\
129	&	12.69692(42)	&	0.089(7)	&	107.4(4.3)	&	10.4	&	$f_{18}-f_{3}$	\\
130	&	0.37694(43)	&	0.088(7)	&	125.8(4.3)	&	10.2	&	$\sim f_{57}$	\\
131	&	0.26211(43)	&	0.087(7)	&	222.9(4.4)	&	10.2	&	$\sim f_{33}$	\\
132	&	16.74591(44)	&	0.086(7)	&	247.3(4.4)	&	10.1	&	$f_{63}-f_{5}$	\\
133	&	0.29352(44)	&	0.086(7)	&	35.7(4.4)	&	10	&	$\sim f_{94}$	\\
134	&	3.09120(44)	&	0.085(7)	&	168.8(4.5)	&	9.9	&	$7f_{\rm orb}$	\\
135	&	31.56027(44)	&	0.085(7)	&	316.8(4.5)	&	9.9	&	$f_{13}$+$f_{63}$	\\
136	&	15.27524(44)	&	0.085(7)	&	134.3(4.5)	&	9.9	&	$2f_{\rm orb}$+$f_{25}$	\\
137	&	26.98604(45)	&	0.084(7)	&	93.1(4.5)	&	9.8	&	$f_{30}$+$f_{5}$	\\
138	&	20.73516(46)	&	0.082(7)	&	321.7(4.6)	&	9.6	&	$f_{37}-f_{61}$	\\
139	&	12.51721(46)	&	0.082(7)	&	17.3(4.7)	&	9.5	&	$f_{1}-f_{13}$	\\
140	&	14.33753(46)	&	0.082(7)	&	278.6(4.6)	&	9.6	&	$\sim f_{6}$	\\
141	&	1.26418(46)	&	0.082(7)	&	357.5(4.7)	&	9.5	&	$f_{8}-f_{36}$	\\
142	&	1.27809(42)	&	0.091(7)	&	130.1(4.2)	&	10.6	&	$\sim f_{141}$	\\
143	&	13.43277(46)	&	0.081(7)	&	354.3(4.7)	&	9.5	&	$\sim f_{11}$	\\
144	&	22.43859(46)	&	0.081(7)	&	69.1(4.7)	&	9.4	&	$f_{20}$+$f_{21}$	\\
145	&	28.57721(48)	&	0.079(7)	&	74.5(4.8)	&	9.2	&	$f_{111}$+$4f_{\rm orb}$	\\
146	&	18.43183(48)	&	0.079(7)	&	310.8(4.8)	&	9.2	&	$f_{16}$+$f_{2}$	\\
147	&	0.95831(48)	&	0.078(7)	&	354.5(4.8)	&	9.1	&	$f_{12}-f_{1}$	\\
148	&	0.80382(48)	&	0.079(7)	&	273.8(4.8)	&	9.2	&	$2f_{83}$	\\
149	&	21.11210(49)	&	0.077(7)	&	42.8(4.9)	&	9	&	$\sim 5f_{\rm orb}$	\\
150	&	20.82322(49)	&	0.078(7)	&	185.8(4.9)	&	9	&	$f_{51}-f_{24}$	\\
151	&	2.21940(49)	&	0.077(7)	&	236.6(4.9)	&	9	&	$2f_{\rm orb}$+$f_{22}$	\\
152	&	1.76059(49)	&	0.077(7)	&	147.5(4.9)	&	9	&	$\sim 4f_{\rm orb}$	\\
153	&	1.78839(49)	&	0.076(7)	&	236.1(5.0)	&	8.9	&	$\sim 4f_{\rm orb}$	\\
154	&	17.38907(51)	&	0.075(7)	&	2.4(5.1)	&	8.7	&	$f_{38}-f_{3}$	\\
155	&	19.13833(50)	&	0.076(7)	&	56.6(5.0)	&	8.9	&	$f_{153}-f_{8}$	\\
156	&	0.71371(51)	&	0.074(7)	&	297.2(5.1)	&	8.6	&	$2f_{57}$	\\
157	&	21.04464(52)	&	0.073(7)	&	317.2(5.2)	&	8.5	&	$f_{4}-2f_{\rm orb}$	\\
158	&	22.92521(52)	&	0.072(7)	&	307.0(5.3)	&	8.4	&	$f_{22}$+$f_{86}$	\\
159	&	13.92248(53)	&	0.072(7)	&	176.2(5.3)	&	8.3	&	$\sim f_{5}$	\\
160	&	1.22299(53)	&	0.072(7)	&	120.9(5.3)	&	8.3	&	$f_{3}-f_{30}$	\\
161	&	19.23617(53)	&	0.071(7)	&	140.3(5.3)	&	8.3	&	$f_{2}$+$f_{61}$	\\
162	&	33.88266(53)	&	0.070(7)	&	351.7(5.4)	&	8.2	&	$f_{\rm orb}$+$f_{85}$	\\
163	&	25.41135(55)	&	0.069(7)	&	105.3(5.5)	&	8	&	$\sim f_{27}$	\\
164	&	0.62256(55)	&	0.069(7)	&	40.4(5.5)	&	8	&	$f_{41}-f_{1}$	\\
165	&	0.69569(54)	&	0.069(7)	&	336.7(5.5)	&	8.1	&	$2f_{110}$	\\
166	&	1.45111(55)	&	0.068(7)	&	308.3(5.6)	&	8	&	$f_{25}-f_{1}$	\\
167	&	1.34760(53)	&	0.070(7)	&	61.4(5.4)	&	8.2	&	$f_{5}-f_{19}$	\\
168	&	1.50620(54)	&	0.070(7)	&	30.6(5.4)	&	8.2	&	$f_{22}-f_{33}$	\\
169	&	25.01279(55)	&	0.068(7)	&	287.8(5.6)	&	7.9	&	$f_{4}$+$f_{96}$	\\
170	&	15.71603(56)	&	0.068(7)	&	193.9(5.6)	&	7.9	&	$f_{13}$+$f_{25}$	\\
171	&	0.22709(57)	&	0.066(7)	&	277.0(5.7)	&	7.7	&	$\sim f_{44}$	\\
172	&	0.24254(57)	&	0.066(7)	&	275.9(5.7)	&	7.7	&	$\sim f_{44}$	\\
173	&	22.53386(57)	&	0.066(7)	&	138.8(5.7)	&	7.7	&	$f_{13}$+$f_{37}$	\\
174	&	3.97175(57)	&	0.066(7)	&	196.9(5.8)	&	7.7	&	$\sim 9f_{\rm orb}$	\\
175	&	0.18641(58)	&	0.065(7)	&	127.5(5.9)	&	7.5	&	$\sim f_{72}$	\\
176	&	0.90475(58)	&	0.065(7)	&	194.6(5.9)	&	7.5	&	$\sim 4f_{\rm orb}$	\\
177	&	27.85423(59)	&	0.064(7)	&	79.1(5.9)	&	7.5	&	$2f_{5}$	\\
178	&	15.94569(59)	&	0.063(7)	&	256.4(6.0)	&	7.4	&	$f_{36}-f_{16}$	\\
179	&	20.49829(59)	&	0.064(7)	&	35.0(6.0)	&	7.4	&	$f_{93}-f_{1}$	\\
180	&	12.55789(59)	&	0.063(7)	&	351.9(6.0)	&	7.4	&	$f_{11}-f_{13}$	\\
181	&	23.00709(59)	&	0.063(7)	&	348.7(6.0)	&	7.4	&	$f_{168}$+$f_{4}$	\\
182	&	34.99184(59)	&	0.063(7)	&	203.3(6.0)	&	7.4	&	$f_{30}$+$f_{51}$	\\
183	&	29.71163(59)	&	0.063(7)	&	285.5(6.0)	&	7.4	&	$f_{41}$+$f_{56}$	\\
184	&	26.02362(60)	&	0.063(7)	&	42.8(6.0)	&	7.3	&	$f_{11}$+$f_{19}$	\\
185	&	23.80061(60)	&	0.062(7)	&	76.0(6.1)	&	7.3	&	$f_{13}$+$f_{158}$	\\
186	&	22.27690(60)	&	0.062(7)	&	52.3(6.1)	&	7.3	&	$f_{174}$+$f_{2}$	\\
187	&	25.73422(61)	&	0.061(7)	&	26.0(6.2)	&	7.2	&	$2f_{69}$	\\
188	&	11.81483(62)	&	0.061(7)	&	126.6(6.3)	&	7.1	&	$f_{129}-f_{13}$	\\
189	&	10.96723(61)	&	0.062(7)	&	82.0(6.1)	&	7.2	&	$f_{29}-f_{25}$	\\
190	&	1.61743(62)	&	0.061(7)	&	33.5(6.3)	&	7.1	&	$2f_{148}$	\\
191	&	2.50159(63)	&	0.060(7)	&	19.3(6.3)	&	7	&	$f_{8}-f_{25}$	\\
192	&	0.64831(63)	&	0.059(7)	&	12.9(6.4)	&	6.9	&	$2f_{91}$	\\
193	&	0.49795(62)	&	0.061(7)	&	282.5(6.2)	&	7.1	&	$\sim f_{42}$	\\
194	&	0.98508(62)	&	0.061(7)	&	24.3(6.2)	&	7.1	&	$2f_{70}$	\\
195	&	14.30148(64)	&	0.059(7)	&	330.7(6.5)	&	6.9	&	$\sim f_{26}$	\\
196	&	17.49824(65)	&	0.058(7)	&	86.7(6.5)	&	6.8	&	$f_{28}$+$f_{8}$	\\
197	&	30.73533(65)	&	0.058(7)	&	218.8(6.5)	&	6.8	&	$f_{1}$+$f_{8}$	\\
198	&	29.46188(65)	&	0.058(7)	&	179.8(6.6)	&	6.7	&	$f_{1}$+$f_{36}$	\\
199	&	1.39601(66)	&	0.057(7)	&	3.6(6.6)	&	6.7	&	$2f_{165}$	\\
200	&	1.88623(65)	&	0.058(7)	&	114.0(6.6)	&	6.7	&	$f_{16}$+$f_{22}$	\\
201	&	1.09631(66)	&	0.057(7)	&	74.7(6.7)	&	6.7	&	$f_{58}-f_{1}$	\\
202	&	2.20756(66)	&	0.057(7)	&	252.3(6.7)	&	6.6	&	$\sim 5f_{\rm orb}$	\\
203	&	22.60543(66)	&	0.057(7)	&	116.6(6.7)	&	6.6	&	$\sim f_{48}$	\\
204	&	1.93103(67)	&	0.057(7)	&	311.8(6.7)	&	6.6	&	$2f_{147}$	\\
205	&	21.99935(68)	&	0.055(7)	&	264.0(6.9)	&	6.4	&	$\sim f_{104}$	\\
206	&	29.10812(68)	&	0.055(7)	&	74.6(6.9)	&	6.4	&	$f_{25}$+$f_{3}$	\\
207	&	20.31343(68)	&	0.055(7)	&	282.7(6.9)	&	6.4	&	$f_{4}-f_{84}$	\\
208	&	2.22918(69)	&	0.055(7)	&	339.1(6.9)	&	6.4	&	$\sim f_{151}$	\\
209	&	35.13345(69)	&	0.055(7)	&	327.4(7.0)	&	6.4	&	$f_{1}$+$f_{95}$	\\
210	&	24.00195(69)	&	0.054(7)	&	326.9(7.0)	&	6.3	&	$f_{191}$+$f_{4}$	\\
211	&	1.77603(69)	&	0.054(7)	&	62.3(7.0)	&	6.3	&	$\sim f_{22}$	\\
212	&	15.52344(69)	&	0.054(7)	&	82.4(7.0)	&	6.3	&	$f_{12}$+$f_{84}$	\\
213	&	33.89347(70)	&	0.054(7)	&	110.6(7.1)	&	6.3	&	$\sim f_{93}$	\\
214	&	0.54223(71)	&	0.053(7)	&	108.2(7.1)	&	6.2	&	$\sim f_{71}$	\\
215	&	20.46327(71)	&	0.053(7)	&	271.2(7.2)	&	6.2	&	$f_{93}-f_{11}$	\\
216	&	0.85120(72)	&	0.053(7)	&	182.8(7.2)	&	6.1	&	$\sim f_{75}$	\\
217	&	0.30948(71)	&	0.053(7)	&	207.2(7.2)	&	6.2	&	$2f_{28}$	\\
218	&	30.60711(72)	&	0.053(7)	&	181.7(7.2)	&	6.1	&	$f_{13}$+$f_{83}$	\\
219	&	18.86541(72)	&	0.052(7)	&	251.9(7.3)	&	6.1	&	$f_{13}$+$f_{54}$	\\
220	&	28.45105(72)	&	0.052(7)	&	294.0(7.3)	&	6.1	&	$f_{145}-f_{16}$	\\
221	&	0.91299(72)	&	0.052(7)	&	314.4(7.3)	&	6.1	&	$\sim 4f_{\rm orb}$	\\
222	&	14.98996(72)	&	0.052(7)	&	84.3(7.3)	&	6.1	&	$f_{156}$+$f_{3}$	\\
223	&	12.54347(73)	&	0.052(7)	&	135.2(7.3)	&	6	&	$\sim f_{180}$	\\
224	&	2.69263(73)	&	0.051(7)	&	268.1(7.4)	&	6	&	$2f_{167}$	\\
225	&	1.40631(73)	&	0.051(7)	&	112.5(7.4)	&	6	&	$\sim f_{199}$	\\
226	&	1.29096(72)	&	0.052(7)	&	215.4(7.3)	&	6.1	&	$\sim f_{142}$	\\
227	&	3.10561(72)	&	0.052(7)	&	115.3(7.3)	&	6.1	&	$\sim 7f_{\rm orb}$	\\
228	&	21.48028(75)	&	0.050(7)	&	275.3(7.5)	&	5.9	&	$\sim f_{4}$	\\
229	&	29.02470(75)	&	0.051(7)	&	158.2(7.5)	&	5.9	&	$f_{113}$+$f_{8}$	\\
230	&	12.24274(73)	&	0.051(7)	&	250.8(7.4)	&	6.0	&	$f_{9}-f_{1}$	\\
231	&	17.77940(75)	&	0.050(7)	&	358.4(7.5)	&	5.9	&	$\sim f_{125}$	\\
232	&	18.21761(75)	&	0.050(7)	&	122.9(7.5)	&	5.9	&	$\sim f_{50}$	\\
233	&	33.91407(76)	&	0.050(7)	&	112.3(7.6)	&	5.8	&	$\sim f_{93}$	\\
234	&	0.52009(76)	&	0.050(7)	&	213.4(7.7)	&	5.8	&	$\sim f_{42}$	\\
235	&	14.01569(76)	&	0.049(7)	&	37.8(7.7)	&	5.7	&	$\sim f_{41}$	\\
236	&	16.48071(77)	&	0.049(7)	&	292.2(7.7)	&	5.7	&	$f_{1}$+$f_{134}$	\\
237	&	20.99109(77)	&	0.049(7)	&	257.3(7.7)	&	5.7	&	$f_{4}-f_{42}$	\\
238	&	35.78743(77)	&	0.049(7)	&	261.8(7.8)	&	5.7	&	$f_{3}$+$f_{4}$	\\
239	&	21.59718(78)	&	0.049(7)	&	346.1(7.8)	&	5.7	&	$f_{16}$+$f_{4}$	\\
240	&	4.10409(78)	&	0.048(7)	&	295.6(7.9)	&	5.6	&	$f_{99}-f_{2}$	\\
241	&	15.60119(78)	&	0.048(7)	&	245.3(7.9)	&	5.6	&	$f_{3}$+$f_{45}$	\\
242	&	35.98053(78)	&	0.048(7)	&	199.2(7.9)	&	5.6	&	$f_{4}$+$f_{58}$	\\
243	&	27.05041(79)	&	0.048(7)	&	92.6(7.9)	&	5.6	&	$2f_{47}$	\\
244	&	23.30678(79)	&	0.048(7)	&	207.7(8.0)	&	5.6	&	$f_{13}$+$f_{99}$	\\
245	&	2.39705(79)	&	0.048(7)	&	170.4(8.0)	&	5.6	&	$f_{18}-f_{52}$	\\
246	&	1.53916(76)	&	0.050(7)	&	77.1(7.6)	&	5.8	&	$f_{22}-f_{44}$	\\
247	&	0.59373(79)	&	0.048(7)	&	302.6(7.9)	&	5.6	&	$2f_{133}$	\\
248	&	0.58291(79)	&	0.048(7)	&	216.5(8.0)	&	5.6	&	$\sim f_{122}$	\\
249	&	15.67380(79)	&	0.048(7)	&	188.2(8.0)	&	5.5	&	$\sim f_{56}$	\\
250	&	15.06411(80)	&	0.047(7)	&	241.7(8.0)	&	5.5	&	$f_{24}$+$f_{5}$	\\
251	&	12.86428(80)	&	0.047(7)	&	4.4(8.0)	&	5.5	&	$\sim f_{69}$	\\
252	&	33.97277(80)	&	0.047(7)	&	85.7(8.1)	&	5.5	&	$f_{12}$+$f_{39}$	\\
253	&	13.02391(81)	&	0.047(7)	&	289.0(8.2)	&	5.4	&	$\sim f_{128}$	\\
254	&	21.13785(81)	&	0.046(7)	&	237.4(8.2)	&	5.4	&	$\sim f_{117}$	\\
255	&	13.93587(81)	&	0.047(7)	&	85.6(8.1)	&	5.4	&	$\sim f_{5}$	\\
256	&	22.83407(77)	&	0.046(6)	&	352.3(7.7)	&	5.4	&		\\
257	&	18.91793(77)	&	0.046(6)	&	50.9(7.8)	&	5.4	&		\\
258	&	2.89397(78)	&	0.046(6)	&	190.7(7.8)	&	5.3	&		\\
259	&	1.81929(78)	&	0.045(6)	&	198.0(7.9)	&	5.3	&		\\
260	&	12.75459(78)	&	0.045(6)	&	228.0(7.9)	&	5.3	&		\\
261	&	2.10714(78)	&	0.045(6)	&	193.6(7.9)	&	5.3	&		\\
262	&	2.84093(79)	&	0.045(6)	&	139.9(7.9)	&	5.3	&		\\
263	&	20.51065(79)	&	0.045(6)	&	324.1(7.9)	&	5.3	&		\\
264	&	17.95705(78)	&	0.045(6)	&	300.7(7.9)	&	5.3	&		\\
265	&	15.56721(79)	&	0.045(6)	&	26.0(7.9)	&	5.3	&		\\
266	&	22.62809(79)	&	0.045(6)	&	76.6(7.9)	&	5.3	&		\\
267	&	0.75336(79)	&	0.045(6)	&	8.5(8.0)	&	5.3	&		\\
268	&	15.99461(79)	&	0.045(6)	&	148.5(8.0)	&	5.2	&		\\
269	&	2.18284(80)	&	0.045(6)	&	316.2(8.0)	&	5.2	&		\\
270	&	3.69883(80)	&	0.044(6)	&	278.1(8.1)	&	5.2	&		\\
271	&	32.84814(80)	&	0.044(6)	&	151.7(8.1)	&	5.1	&		\\
272	&	31.71475(81)	&	0.044(6)	&	257.8(8.2)	&	5.1	&		\\
273	&	0.09114(81)	&	0.044(6)	&	182.0(8.2)	&	5.1	&		\\
274	&	12.26746(81)	&	0.044(6)	&	319.2(8.2)	&	5.1	&		\\
275	&	1.36666(81)	&	0.044(6)	&	300.2(8.2)	&	5.1	&		\\
276	&	1.05872(81)	&	0.044(6)	&	153.8(8.2)	&	5.1	&		\\
277	&	16.87773(82)	&	0.044(6)	&	124.3(8.2)	&	5.1	&		\\
278	&	31.44080(82)	&	0.043(6)	&	98.1(8.3)	&	5.1	&		\\
\end{longtable}
\end{center}

\section{Sidelobes}

This appendix includes the results for the potential rotational splitting of frequencies (sidelobes). For this, the equation of Ledoux (1951) (\textit{c.f.} Aerts \textit{et al.} 2010) is used:
\begin{equation}
f=f_0 + m(1-C_{nl})f_{\rm rot},
\end{equation}
where $f_0$ is the central detected frequency, $m=\pm l$ the azimuthal order (up to $l=2$, where $l$ is the spherical degree), $C_{nl}=0$ (\textit{i.e.} Ledoux constant; assumed as zero for simplicity), and $f_{\rm rot}=f_{\rm orb}=0.441604$~d$^{-1}$ (synchronous rotation). In Table~5, the columns 1-3 contain the potential combination based on the aforementioned equation and its error. The columns 4-6 contain the increasing number ($i$) of the detected frequency, its value ($f$) and its error. The 7th column includes the sum of the errors (columns 3 and 6). The 8th column includes the absolute difference between the calculated combination (column 2) and the detected frequency (column 5). The last column contains the check whether this difference is larger than the summary of the errors of the detected and the calculated frequencies. If yes, then that means that the detected frequency has such a value that cannot be attributed as possible spilt frequency of another central one. If no, then the detected frequency indeed can be considered as a split frequency. Empty cells denote that no values close to the combination frequencies were found among the list of detected frequencies.

\MakeTable{ccc| ccc| c| c| c}{14cm}{Results for possible sidelobes of KIC~6629588.}
{\hline	
										
\multicolumn{3}{p{4cm}}{potential combination}					&	\multicolumn{3}{p{4cm}}{detection}	&	sum error	&	Abs. Diff.	&	Diff$>$error	\\
\hline																	
	&		&	error	&	$i$	&	$f$	&	error	&		&		&		\\
\hline																	
\multicolumn{9}{p{13cm}}{For $m=\pm 2$}																\\
\hline																	
$f_1+f_{\rm orb}$	&	13.83814	&	0.00001	&	$f_{67}$	&	13.83855	&	0.00022	&	0.00023	&	0.00041	&	YES	\\
$f_1-f_{\rm orb}$	&	12.95493	&	0.00001	&	-	&	-	&	-	&		&		&		\\
$f_2+f_{\rm orb}$	&	18.75762	&	0.00002	&	-	&	-	&	-	&		&		&		\\
$f_2-f_{\rm orb}$	&	17.87441	&	0.00002	&	-	&	-	&	-	&		&		&		\\
$f_4+f_{\rm orb}$	&	21.93491	&	0.00003	&	$f_{51}$	&	21.94785	&	0.00017	&	0.0002	&	0.01294	&	YES	\\
$f_4-f_{\rm orb}$	&	21.0517	&	0.00003	&	$f_{157}$	&	21.04464	&	0.00052	&	0.00055	&	0.00706	&	YES	\\
$f_5+f_{\rm orb}$	&	14.37222	&	0.00004	&	-	&	-	&	-	&		&		&		\\
$f_5-f_{\rm orb}$	&	13.48901	&	0.00004	&	-	&	-	&	-	&		&		&		\\
$f_6+f_{\rm orb}$	&	14.76729	&	0.00004	&	-	&	-	&	-	&		&		&		\\
$f_6-f_{\rm orb}$	&	13.88408	&	0.00004	&	-	&	-	&	-	&		&		&		\\
$f_7+f_{\rm orb}$	&	13.9027	&	0.00005	&	$f_{81}$	&	13.91321	&	0.00028	&	0.00032	&	0.01051	&	YES	\\
$f_7-f_{\rm orb}$	&	13.01949	&	0.00005	&	$f_{253}$	&	13.02391	&	0.00081	&	0.00086	&	0.00442	&	YES	\\
$f_8+f_{\rm orb}$	&	17.77712	&	0.00004	&	$f_{231}$	&	17.7794	&	0.00075	&	0.00079	&	0.00228	&	YES	\\
$f_8-f_{\rm orb}$	&	16.89391	&	0.00004	&	-	&	-	&	-	&		&		&		\\
\hline																	
\multicolumn{9}{p{13cm}}{For $m=\pm 2$}																	\\
\hline																	
$f_1+2f_{\rm orb}$	&	14.27974	&	0.00001	&	$f_{3}$	&	14.2799	&	0.00002	&	0.00003	&	0.00016	&	YES	\\
$f_1-2f_{\rm orb}$	&	12.51332	&	0.00001	&	$f_{139}$	&	12.51721	&	0.00046	&	0.00047	&	0.00388	&	YES	\\
$f_2+2f_{\rm orb}$	&	19.19922	&	0.00002	&	-	&	-	&	-	&		&		&		\\
$f_2-2f_{\rm orb}$	&	17.43281	&	0.00002	&	$f_{15}$	&	17.43284	&	0.00007	&	0.00009	&	0.00004	&	NO	\\
$f_4+2f_{\rm orb}$	&	22.37651	&	0.00003	&	-	&	-	&	-	&		&		&		\\
$f_4-2f_{\rm orb}$	&	20.6101	&	0.00003	&	$f_{263}$	&	20.51065	&	0.00079	&	0.00082	&	0.09945	&	YES	\\
$f_5+2f_{\rm orb}$	&	14.81382	&	0.00004	&	$f_{25}$	&	14.83753	&	0.00011	&	0.00015	&	0.02371	&	YES	\\
$f_5-2f_{\rm orb}$	&	13.04741	&	0.00004	&	$f_{30}$	&	13.04554	&	0.00011	&	0.00015	&	0.00187	&	YES	\\
$f_6+2f_{\rm orb}$	&	15.20889	&	0.00004	&	-	&	-	&	-	&		&		&		\\
$f_6-2f_{\rm orb}$	&	13.44247	&	0.00004	&	$f_{11}$	&	13.44204	&	0.00005	&	0.0001	&	0.00043	&	YES	\\
$f_7+2f_{\rm orb}$	&	14.3443	&	0.00005	&	$f_{12}$	&	14.34577	&	0.00006	&	0.00011	&	0.00146	&	YES	\\
$f_7-2f_{\rm orb}$	&	12.57789	&	0.00005	&	$f_{19}$	&	12.58054	&	0.00007	&	0.00012	&	0.00266	&	YES	\\
$f_8+2f_{\rm orb}$	&	18.21873	&	0.00004	&	$f_{50}$	&	18.21864	&	0.00017	&	0.00021	&	0.00008	&	NO	\\
$f_8-2f_{\rm orb}$	&	16.45231	&	0.00004	&	$f_{85}$	&	16.44982	&	0.00028	&	0.00032	&	0.00249	&	YES	\\
\hline																	
}											

\end{appendix}

\end{document}